%Paper: gr-qc/9211002
%From: "Jonathan Z. Simon" <JSIMON%umdhep.BITNET@VTVM2.CC.VT.EDU>
%Date: Tue, 3 Nov 92 12:14 EDT

% TeX file begins here. Delete all lines above this one before TeXing.
%
% There is a single postscript file attached at the end of this TeX file,
% beginning after the final \end of the Tex file. That postscript
% file, if sent to a postscript printer, will result in 4 pages of figures.
% The postscript file need not be stripped off to TeX the paper.

\overfullrule 0pt
\font\bigbf = cmbx10 scaled \magstep2
\def\d{\delta}
\def\m{\noalign{\medskip}}
\def\n{\nabla}

\def\n{\nabla}
\def\p{\partial}
\baselineskip 20pt
%%%%%%%%%%%%%%%%%%%%%%%%%%%%%%%%%%%%%%%%%%%%%%%%%%%%%%%%%%%%%%%%%%%%%%%%
%
%   Here come chapter, section, subsection & appendix macros.
%
\newcount\chapternumber      \chapternumber=0
\newcount\sectionnumber      \sectionnumber=0
\newcount\equanumber         \equanumber=0
\let\chapterlabel=\relax
\let\sectionlabel=\relax
\newtoks\chapterstyle        \chapterstyle={\Number}
\newtoks\sectionstyle        \sectionstyle={\chapterlabel\Number}
\newskip\chapterskip         \chapterskip=\bigskipamount
\newskip\sectionskip         \sectionskip=\medskipamount
\newskip\headskip            \headskip=8pt plus 3pt minus 3pt
\newdimen\chapterminspace    \chapterminspace=15pc
\newdimen\sectionminspace    \sectionminspace=10pc
\newdimen\referenceminspace  \referenceminspace=25pc
\def\chapterreset{\global\advance\chapternumber by 1
   \ifnum\equanumber<0 \else\global\equanumber=0\fi
   \sectionnumber=0 \makechapterlabel}
\def\makechapterlabel{\let\sectionlabel=\relax
   \xdef\chapterlabel{\the\chapterstyle{\the\chapternumber}.}}
\def\alphabetic#1{\count255='140 \advance\count255 by #1\char\count255}
\def\Alphabetic#1{\count255='100 \advance\count255 by #1\char\count255}
\def\Roman#1{\uppercase\expandafter{\romannumeral #1}}

\def\Number#1{\number #1}
\def\BLANC#1{}
\def\chapter#1{\par \penalty-300 \vskip\chapterskip
   \spacecheck\chapterminspace
   \chapterreset \titlestyle{\chapterlabel\ #1}
   \TableOfContentEntry c\chapterlabel{#1}
   \nobreak\vskip\headskip \penalty 30000
   \wlog{\string\chapter\space \chapterlabel} }

\def\section#1{\par \ifnum\the\lastpenalty=30000\else
   \penalty-200\vskip\sectionskip \spacecheck\sectionminspace\fi
   \global\advance\sectionnumber by 1
   \xdef\sectionlabel{\the\sectionstyle\the\sectionnumber}
   \wlog{\string\section\space \sectionlabel}
   \TableOfContentEntry s\sectionlabel{#1}
   \noindent {\caps\enspace\sectionlabel\quad #1}\par
   \nobreak\vskip\headskip \penalty 30000 }

\def\unnumberedchapters{\let\makechapterlabel=\relax \let\chapterlabel=\relax
   \sectionstyle={\BLANC}\let\sectionlabel=\relax \sequentialequations }
%
%%%%%%%%%%%%%%%%%%%%%%%%%%%%%%%%%%%%%%%%%%%%%%%%%%%%%%%%%%%%%%%%%%%%%%%%
%
%   Here come macros for equation numbering.
%
\def\eqname#1{\relax \ifnum\equanumber<0
     \xdef#1{{\noexpand\rm(\number-\equanumber)}}%
       \global\advance\equanumber by -1
    \else \global\advance\equanumber by 1
      \xdef#1{{\noexpand\rm(\chapterlabel\number\equanumber)}} \fi #1}

\def\eqn{\eqno\eqname}

\def\eqinsert#1{\noalign{\dimen@=\prevdepth \nointerlineskip
   \setbox0=\hbox to\displaywidth{\hfil #1}
   \vbox to 0pt{\kern 0.5\baselineskip\hbox{$\!\box0\!$}\vss}
   \prevdepth=\dimen@}}

\magnification=1200
\def\pmb#1{\setbox0=\hbox{#1}%
\kern-.025em\copy0\kern-\wd0
\kern.05em\copy0\kern-\wd0
\kern-.025em\raise.0433em\box0}

\chapternumber=1
\makechapterlabel
\def\la{\langle}
\def\ra{\rangle}
\def\n{\nabla}
\def\a{\alpha}
\def\b{\beta}
\def\sqr#1#2{{\vcenter{\vbox{\hrule height.#2pt
        \hbox{\vrule width.#2pt height#1pt \kern#1pt
          \vrule width.#2pt}
        \hrule height.#2pt}}}}
\def\square{\mathchoice\sqr54\sqr54\sqr33\sqr23}
\raggedbottom
%Content-Length: 89109

\rightline{WISC-MILW-92-TH-14}
\bigskip
\centerline{\bigbf{Einstein Equation with Quantum Corrections}}
\centerline{\bigbf{Reduced to Second Order}}
\bigskip
\centerline{Leonard Parker\footnote * {Electronic Mail:
leonard@cosmos.phys.uwm.edu}}
\centerline{Jonathan Z. Simon\footnote \dag {Electronic Mail:
jsimon@csd4.csd.uwm.edu}\footnote \ddag {Current Address:
Dept. of Physics, U. of Maryland, College Park, MD
20742, U.S.A.}}
\bigskip
\centerline{\it{Department of Physics}}
\centerline{\it{University of Wisconsin-Milwaukee}}
\centerline{\it{Milwaukee, WI  53201}}
\centerline{\it{U.S.A.}}
\bigskip \bigskip

\midinsert\narrower{\narrower{We consider the Einstein equation with first
order (semiclassical) quantum corrections. Although the quantum
corrections contain up to fourth order derivatives of the metric, the
solutions which are physically relevant satisfy a reduced equations which
contain derivatives no higher than second order.  We obtain the reduced
equations for a range of stress-energy tensors.  These reduced equations
are suitable for numerical solution, are expected to contain fewer
numerical instabilities than the original fourth order equations, and
yield only physically relevant solutions.  We give analytic and numerical
solutions or reduced equations for particular examples, including
Friedmann-Lema\^\i tre
universes with cosmological constant, a spherical body of constant
density, and more general conformally flat metrics.}}

\endinsert

\vfil\eject

\centerline{\bigbf{1.  Introduction}}

Quantum corrections to general relativity are expected to be important in
regimes where the curvature is near the Planck scale
$(l_{pl}={\sqrt{G\hbar /c^3}} \approx 1.6\times 10^{-33}$ cm).  In a
regime where the curvature approaches but always remains (significantly)
less than the Planck scale, a semiclassical approximation to the full
theory of quantum gravity should be sufficient.  Examples of this regime
include small evaporating black holes, when still much larger than the
Planck mass $(m_{pl} = {\sqrt{\hbar c/G}} \approx 2.2\times 10^{-5}$ g),
and the early universe after it has reached a size of many Planck lengths.
 In the standard semiclassical approximation, the gravitational field
itself is treated classically, but is driven by the expectation value of
quantum matter stress-energy.

The form of the semiclassical corrections to Einstein's field
equations is known for many important cases.$^1$  For example,
for conformally flat classical backgrounds (in four dimensions),
when the quantum state is constructed from the conformal vacuum,
the corrections are completely determined by local geometry (the
metric, the curvature, the covariant derivatives of the curvature):$^2$
$$\eqalign{\kappa\langle T_{ab}\ra &= R_{ab} - {1\over 2}
Rg_{ab}+\Lambda g_{ab}\cr
&\quad +\a_1\hbar \left( {1\over 2} R^2g_{ab}-2RR_{ab}-
2\square Rg_{ab} +2\n_a\n_b R\right)\cr
&\quad +\a_2\hbar \left( {1\over 2} R^{cd}R_{cd}g_{ab}-
\square R_{ab} - {1\over 2} \square Rg_{ab}+\n_a\n_b R-
R^{cd}R_{cadb}\right)\cr
&\quad +\a_3\hbar \left( - {1\over{12}}
R^2g_{ab}+R^{cd}R_{cadb}\right) + O(\hbar^2).\cr} \eqn\one$$
The parameters $\a_1$, $\a_2$, and $\a_3$ depend on the particular
form of matter and regularization scheme, so we do not assume
specific values or signs.  Factors of $\hbar$ have been made
explicit.  Because the corrections are purely geometric, it is
common to consider them not as matter source terms but as metric
field terms (despite their matter origin). Non-conformally flat
backgrounds can have more quantum corrections than Eq.\ \one .
Of the examples above,
the state-independent terms of Eq.\ \one\ do not contribute in the
case of the black hole, where the exterior Ricci curvature vanishes, but
they do contribute in the case of cosmological solutions.
Several cosmological models are examined
below.  Because the new terms contain fourth derivatives in the
metric, the new terms qualitatively change the field equations from
system of second order equations to a system of fourth order
equations.

The new fourth order theory contains whole new classes of solutions
unavailable to the classical theory.  Many of these solutions have
been examined.$^3$  One set of these solutions is particularly
disturbing however.  Solutions to the linearized theory around a
flat background strongly indicate that flat space is unstable to
ultraviolet fluctuations.$^{4,5}$  Using a $1/N$ approximation,
Hartle and Horowitz showed that the ultraviolet instability can be
made to occur at frequencies arbitrarily far below the Planck
frequency, indicating that the instabilities cannot be easily fixed
by calling the full quantum theory of gravity to the rescue.$^5$
Additional instabilities have also been found by Suen.$^{20}$
This strongly indicates that semiclassical gravity, if all its
solutions are considered physical, is not a good description of the
near classical limit of quantum gravity.

It was shown in earlier work$^6$ that it is possible, and indeed
desirable, to modify semiclassical gravity in a way suggested by
and consistent with the perturbative nature of its derivation.  The
effective action and field equations of semiclassical gravity are
perturbative expansions (formally, asymptotic expansions) in powers
of $\hbar$, truncated at first order in $\hbar$.  All behavior
higher order and non-perturbative in $\hbar$ has already been lost
in the process of deriving the (approximate) effective action and
field equations.  Self-consistency then requires that only the
solutions that are also asymptotic expansions in powers of $\hbar$,
truncated to first order, will be approximations to solutions of
the full, non-perturbative effective action.  Solutions not in this
form are likely to be unphysical and should be excluded.  A simple
model, presented below, will demonstrate that retaining
non-perturbative solutions to a perturbatively derived higher derivative
action
results in false predictions.  The non-perturbatively expandable
solutions are spurious artifacts arising from the higher
derivatives appearing in the perturbative correction, and will be
referred to as spurious.  For convenience, perturbatively
expandable solutions will sometimes be referred to as physical,
since only they correspond to predictions of the self-consistent
semiclassical theory.  For semiclassical gravity, it has been shown
that the physical solutions show no signs of any instability of
flat space (to first order in $\hbar$).$^6$

The easiest way of implementing the self-consistent method in
semiclassical gravity is by reducing the fourth order equation, which has
both physical and non-physical solutions, to a second order equation,
which has only physical solutions (with one caveat described below).  This
iterative reduction has been demonstrated in a similar context by Bel and
Sirousse-Zia for the case $\a_3=0.^7$  Much of the reduction (though not
always all) can be done covariantly.  It is clearly more efficient to find
solutions to the reduced second order equations, almost all of which are
physical, rather than finding all solutions to the full fourth order
equations, most of which are spurious, and only using those which are
physical.

The aim of this work is to apply the reduction of order to a wide
variety of gravitational systems.  These include computing the
reduced semiclassical equations for Friedmann cosmologies
(homogeneous isotropic solutions with perfect fluid matter),
Friedmann-Lema\^\i tre cosmologies (Friedmann cosmologies
with cosmological constant), an interior Schwarzschild
solution, and the general, conformally flat metric
in terms of its conformal factor.
Examples of analytic and numerical methods are employed.
In particular, we find the exact semiclassical solutions for
spatially flat, radiation-filled Friedmann cosmologies, and exact
and numerical semiclassical ``bounce'' solutions for
radiation-filled Friedmann-Lema\^\i tre cosmologies.  As expected,
the semiclassical corrections usually play only a small role in
most systems far from the Planck scale.  There are exceptions to
this rule of thumb, however, which we demonstrate by analyzing the
semiclassical corrections to the (unstable) eternal Einstein
universe. Here the corrections can cause large
deviations from the classical solutions and yet remain within the
domain of reliability.

We do not explicitly account for
effects of particle creation (except in conformally flat spacetimes),
only for ``state-independent''
contributions to the stress energy.
\bigskip
\bigskip
\vfil\eject

\chapternumber=2
\makechapterlabel
\equanumber=0

\centerline{\bigbf{2.  Review of Semiclassical Corrections}}

The semiclassical field equations of general relativity (including
cosmological constant) take the form
$$R_{ab}- {1\over 2} Rg_{ab}+\Lambda g_{ab}=\kappa \la
T_{ab}\ra \eqn\rone$$
where $\la T_{ab}\ra =O(\hbar )$ is the expectation value or
transition amplitude of the matter stress-energy tensor.  For
convenience, we consider only massless, conformally coupled fields
(of arbitrary spin).  We may reasonably restrict the form of $\la
T_{ab}\ra$ to obey Wald's physical axioms:$^8$
\item{1)}  covariant conservation
\item{2)}  causality
\item{3)}  standard results for `off-diagonal' matrix elements
\item{4)}  standard results in Minkowski space.

\noindent Wald showed that any $\la T_{ab}\ra$ that obeys the
first 3 axioms is unique up to the addition of a local, conserved
tensor.  Furthermore, any local, conserved tensor can reasonably be
considered part of the geometrical dynamics and so be written on
the left-hand side of the field equations.  We shall do so,
rewriting Eq.\ \rone\ as:
$$R_{ab} - {1\over 2} Rg_{ab} +\Lambda g_{ab} +
\Omega_{ab} = \kappa\la T_{ab}\ra \eqn\rtwo$$
where $\Omega_{ab}$ is conserved and purely local, i.e. it is
constructed purely from the metric, the curvature, and a (finite
number of) its covariant derivatives.

Only terms in $\Omega_{ab}$ that are first order in $\hbar$
will be considered, since the semiclassical approximation already
neglects higher order contributions.$^{21}$  Any term contributing to
$\Omega_{ab}$ with a constant coefficient proportional to
$\hbar$ must have dimensions of $(length)^{-4}$, since the only
length scale is the Planck length, $l_{pl}$, and $\hbar = l^2_{pl}$
in units where $G=1$.  This restricts the form of $\Omega_{ab}$
for general spacetimes in four dimensions to linear combinations of
two possible contributing terms.$^9$
$$\eqalign{^{(1)}H_{ab} &= {1\over{\sqrt g}}
{\delta\over{\delta g^{ab}}} \int d^4x \sqrt g R^2\cr
&= {1\over 2} R^2g_{ab} - 2RR_{ab} - 2\square Rg_{ab}
+ 2\n_a\n_b R\cr} \eqn\rthree$$
$$\eqalign{^{(2)}H_{ab} &= {1\over{\sqrt g}} {\delta\over{\delta
g^{ab}}} \int d^4x \sqrt g R^{cd}R_{cd}\cr
&= {1\over 2} R^{cd}R_{cd}g_{ab}-\square R_{ab} -
{1\over 2} Rg_{ab} + \n_a\n_b R-R^{cd}R_{cadb}\cr} \eqn\rfour$$
These two expressions are automatically conserved from their variational
definitions.  They are also fourth order in time derivatives of the
metric.  In conformally flat, four-dimensional spacetimes (where
the Weyl tensor, $C^c_{def}$, vanishes),
$^{(1)}H_{ab}$ and $^{(2)}H_{ab}$ no longer remain linearly
independent (in this case $^{(1)}H_{ab}=3\;
^{(2)}H_{ab}$).  However, a new quantity appears,
$$\eqalign{^{(3)}H_{ab} &= - {1\over{12}} R^2g_{ab} +
R^{cd}R_{cadb}\cr
&= -R_a^{\;c}R_{cb}+{\textstyle{2 \over 3}}RR_{ab}+
{\textstyle{1 \over 2}}R_{cd}R^{cd}g_{ab}
-{\textstyle{1 \over 4}}R^2g_{ab},\cr} \eqn\rfive$$
which is conserved only in conformally flat spacetimes, but not as
a result of a variational derivation, nor as the limit of a
conserved quantity in non-conformally flat space-times.$^{10}$  It
is second order in derivatives of the metric, unlike
$^{(1)}H_{ab}$ and $^{(2)}H_{ab}$.  Nevertheless, it is
allowed by Wald's axioms, and, in general, contributes to the
conformal anomaly.  The most general expression for
$\Omega_{ab}$ is then
$$\Omega_{ab} = \a_1\hbar\; ^{(1)}H_{ab} + \a_2\hbar\;
^{(2)}H_{ab} + \a_3\hbar\; ^{(3)}H_{ab} + O(\hbar^2) ,
\eqn\rsix$$
where it should be understood that the $^{(3)}H_{ab}$ term is
only present when $^{(1)}H_{ab} = 3\; ^{(2)}H_{ab}$.
Values of $\a_1$, $\a_2$, and $\a_3$ are predicted by specific
matter couplings and regularization schemes, but we will treat them
as free parameters.  Factors of $\hbar$ have been made explicit.
Inserting Eqs.\ \rthree -\rsix\ into Eq.\ \rtwo\ produces Eq.\ \one
{}.

As pointed out above, the new fourth order theory contains whole
new classes of solutions unavailable to the classical, second order
theory.  It is the new solutions that would indicate the
instability of flat space.  Since flat space is experimentally
stable (or at least very metastable), this strongly indicates that
semiclassical gravity, if all its solutions are considered
physical, is not a good description of the near classical limit of
quantum gravity.  However, the derivation of the theory is well
founded, and it seems likely that some of the solutions do
correspond to what we expect from quantum corrections to classical
theory.  It is necessary to break up the solutions to the
semiclassical field equations into those we do not consider part of
the theory (spurious or ``unphysical'') and the rest of the solutions
(``physical''), which contain all the important information
of the theory.

That a theory should contain unphysical solutions should not be
surprising to anyone who has examined Dirac's theory of charged
particles including electromagnetic back-reaction.$^{11}$  The
electromagnetic back-reaction problem shares several features with
the quantum back-reaction described above.  Its most important
features are:
\item{1)}  A small correction term to the equations of motion
changes the order of the equations of motion (from second order to
third order)
\item{2)}  Not all solutions to the new equations of motion are
physical; some must be excluded by external criteria.

\noindent Dirac's equation of motion is
$$\ddot x^\mu - {2\over 3} {e^2\over m} ( {\buildrel ...\over
x}^\mu - \ddot x_\nu\ddot x^\nu \dot x^\mu ) = {e\over m}
F_\nu{}^\mu \dot x^\nu . \eqn\rseven$$
The classic example of a non-physical solution has $F_\nu{}^\mu
= 0$ but exponentially increasing acceleration,
$$\dot x^\mu = [\cosh (\exp \left( {3mc^2\over{2e^2}}\tau\right)
),\; \sinh (\exp \left( {3mc^2\over{2e^2}}\tau\right) ), \; 0,\; 0]
. \eqn\reight$$
If the theory is to make useful predictions, unphysical solutions
such as this must be excluded.  In fact this so-called ``runaway''
solution shares much in common with the particular solutions to
linearized semiclassical gravity which contribute to the
instability of flat space.  Other theories with higher derivative
corrections, such as cosmic strings with rigidity corrections, can
give unphysical solutions with negative kinetic energy.$^{12-14}$

Similar problems occur even for semiclassical Quantum
Electrodynamics (QED).  The running of the electron charge coupling
results in an effective action and Lagrangian density with
higher order corrections$^{15}$
$${\cal L}_{\rm{eff}} = - {1\over 4} F^{\mu\nu} \left( 1-
{\hbar e^2_L\over{60\pi^2m^2}}\square\right) F_{\mu\nu} +\ldots +
{\hbox{matter ,}} \eqn\rnine$$
where $e_L$ is the low energy electron charge.  In this case it is clear
what should be done.  The higher derivatives in Eq.\ \rnine\ do not
correspond to new degrees of freedom for the electromagnetic field, but
rather to the running of the charge.  Since the effective action is a
perturbative expansion in  $\hbar$, one may first solve the lower order
equations of motion, and then solve higher orders iteratively.  Treating
Eq.\  \rnine\ as a true fourth order equation would result in a theory very
different from classical electrodynamics, where the electromagnetic field
possesses negative energy modes. This is clearly undesirable from a
physical point of view (though the resulting theory is mathematically well
defined).

One technique of distinguishing and excluding unphysical solutions of
higher derivative theories, is called the self-consistent
method,$^{13,16}$ and was first applied in the case of the classical Dirac
electron by Bhabha.$^{17}$  In the case of semiclassical QED shown above,
it is equivalent to the obvious method of constructing higher order
solutions iteratively.  It can be applied to any theory for which higher
derivative terms in the field equations are perturbative corrections to a
lower order theory, and is most naturally applied to theories derived from
an effective action with a small parameter that has been expanded in a
power series.  One expects extrema of that expanded action to be
(perturbative) approximations to the extrema of the full effective action.
 For ordinary actions, whose variations give second order differential
equations, this is usually true.  For actions which possess higher
derivative expansions, however, the opposite is true:  most solutions of
the perturbatively expanded field equations {\it{do not even have}} a
perturbative expansion ({\it{i.e.}} they are not analytic in the expansion
parameter as the parameter approaches zero).

An obvious cure for this behavior is to only rely  on those
solutions that {\it{are}} perturbatively expandable (in the same
sense as their effective action) to be physical.  All other
solutions are treated as spurious by-products of the higher
derivatives, not to be considered part of the theory (e.g. the
runaway Dirac electron).

The self-consistent approach is extremely powerful.  It removes all
runaway and negative energy solutions.  It can be applied to semiclassical
gravity as easily as to Dirac's classical electron.  The amount of initial
data required to specify a physical solution is the same as for the
original uncorrected theory.$^{18}$  In the case of cosmic strings, where
the full action is known exactly, any method not equivalent to the
self-consistent approach simply gives the wrong results.  The
self-consistent approach seems clearly applicable to the case of
semiclassical gravity.

The solutions of semiclassical gravity are solutions to a fourth
order differential equation.  The amount of initial data required
to specify a {\it{physical solution}}, however, is the same as for classical
gravity, which is given by solutions to a second order differential
equation.  One can often find a second order differential
equation which contains all the physical solutions to semiclassical
gravity.  This reduction of order greatly simplifies the process of
finding physical solutions since most unphysical solutions are
completely bypassed.  The reduction is performed iteratively, using
lowest (perturbative) order results to simplify the higher order
semiclassical corrections.  Reductions for corrections containing
$^{(1)}H_{ab}$ and $^{(2)}H_{ab}$ have been calculated for
several cases by Bel and Sirousse-Zia.$^7$  The non-linearity of
general relativity makes reduction of order awkward for the most
general solutions unless the stress-energy tensor has an extremely
simple dependence on the metric.  Just as in the case of classical
gravity, however, the presence of symmetries can make soluble an
otherwise intractable problem.  We begin with the example of
semiclassical corrections in Friedmann-Lema\^\i tre
universes.

There is a small but important caveat regarding removing spurious
solutions by the reduction of order.  All the physical solutions to the
higher order field equations are also solutions to the reduced equations,
but,  still, not all solutions to the reduced equations may be physical
solutions.  If the corrections are nonlinear in the field variable, there
may remain a much smaller number of spurious that must still be expunged.
On reduction, however, it is often easy to identify the unphysical
solutions and remove them.

Reduction of order (and the above caveat) are well demonstrated by a simple
example. The example uses a toy ``semiclassical'' equation
$$\ddot x = -\omega^2x + \a\hbar {\buildrel ....\over x}^2 +
O(\hbar^2) , \eqn\rten$$
which is non-linear in the fourth derivative correction.  This is
straightforwardly reduced with the substitution
$$\hbar {\buildrel ....\over x} = -\hbar \omega^2\ddot x +
O(\hbar^2) , \eqn\releven$$
resulting in the reduced equation
$$\ddot x = -\omega^2x+\alpha\hbar\omega^4\ddot x^2+O(\hbar^2). \eqn\rtwelve$$
This second order equation still contains all the physical solutions
(perturbatively expandable in $\hbar$), but because of the quadratic nature of
the equation, contains solutions non-perturbative in $\hbar$.  Solving the
quadratic equation gives
$$\ddot x = \cases{ {-2\omega^2x\over{1+{\sqrt{1+4\alpha\hbar\omega^6x}}}} = -
\omega^2x+\alpha\hbar\omega^4x^2+O(\hbar^2)\cr\m
{1+{\sqrt{1+4\alpha\hbar\omega^6x}}\over{2\a\hbar\omega^4}} =
{1\over{\a\hbar\omega^4}} + \omega^2x + \ldots\cr} . \eqn\rthteen$$
The first of these manifestly produces only solutions perturbative in $\hbar$
and
so contains only physical solutions.  The second set is found only after
treating
$\hbar$ non-perturbatively:  it would not be found using strictly perturbative
methods and thus describes non-physical solutions.  The point is that that even
though Eq.\ \rtwelve\ is reduced to second order, some care must still be
exercised until the equation has been put into form of the first line of Eq.\
\rthteen .  In practice we will often use reduced equations in the form of Eq.\
\rtwelve , but if numerical methods are used, the form of Eq.\ \rtwelve\
may not be adequate and one may need analogs of Eq.\ \rthteen .
\bigskip
\bigskip
\vfil\eject

\chapternumber=3
\makechapterlabel
\equanumber=0

\centerline{\bigbf{3.  Friedmann-Lema\^\i tre Models}}
\centerline{\bigbf{With Quantum Corrections}}

In this section, we consider general Robertson-Walker
metrics of the form
$$ds^2 = -dt^2 + a(t)^2 \left( {{dr^2} \over {1-kr^2}}
           + r^2 d\theta^2 + r^2 \sin^2 \theta d\phi^2 \right), \eqn\RWmetric$$
where $k$ takes the values $\pm 1$ or $0$.
The matter consists of radiation and the cosmological constant need not be
zero.  The $(t,t)$ component of the Einstein equation with first order
quantum corrections is of third-order in time derivatives of the scale
factor $a(t)$. We first reduce the equation to first-order (the same order
as the corresponding classical Einstein equation), which the physically
relevant solutions satisfy. For various values of the spatial curvature
$k$ and cosmological constant $\Lambda$, we obtain analytic and numerical
solutions. For some solutions within  this class of Friedmann-Lema\^\i tre
universes we find that there are models for which the first order quantum
corrections  remain small at all times. In some cases, the effect of small
quantum corrections can cause a large deviation from the corresponding
classical solution over a long period of time.

In this conformally flat class of metrics,
the Einstein equations with quantum corrections are of the form,
$$\eqalign{R_{ab} &- (1/2) g_{ab} R + \Lambda g_{ab}\cr
 &+ \hbar \alpha_1\, {}^{(1)} H_{ab} +
     \hbar \alpha_3\, {}^{(3)} H_{ab} + O(\hbar^2) = \kappa T_{ab}\cr}
\eqn\Einstein$$
Here $\Lambda$ is the cosmological constant, and
$T_{ab}$ includes classical matter contributions and
the lowest order expectation value of quantum matter,
and ${}^{(1)} H_{ab}$ and ${}^{(3)} H_{ab}$ are defined
bye Eqs.\ \rthree and \rfive. The
state-independent local quantum corrections of order $\hbar$
are included in the $H$ terms on the left-hand-side.

With the metric of Eq.\ \RWmetric ,
$$\eqalign{^{(1)} H_{tt} &= {{-18 {k^2}}\over {{{a}^4}}}
     + {{36 k {{\dot a}^2}}\over {{{a}^4}}}
     + {{54 {{\dot a}^4}}\over {{{a}^4}}} - {{36 {{\dot a}^2} \ddot a}\over
{{{a}^3}}}
     + {{18 {{\ddot a}^2}}\over {{{a}^2}}}
     - {{36 \dot a a^{(3)}}\over {{{a}^2}}},\cr} \eqn\Honett$$
and
$$^{(3)} H_{tt} =
  {{3 {k^2}}\over {{{a}^4}}}
  + {{6 k {{\dot a}^2}}\over {{{a}^4}}}
  + {{3 {{\dot a}^4}}\over {{{a}^4}}}. \eqn\Hthreett$$

The $(t,t)$ component of the generalized
Einstein equation\ \Einstein , with matter
consisting of radiation, is
$$\eqalign{0\;  = \; &- \Lambda - \kappa \rho_0 a_0^4 / a^4
 + (3 k + 3 \dot a^2)/ a^2\cr
&+ \alpha_1 \hbar
 ( -18 {k^2} + 36 k {{\dot a}^2} + 54 {{\dot a}^4}
 - 36 a {{\dot a}^2} \ddot a
 + 18 {{a}^2} {{\ddot a}^2}
 - 36 {{a}^2} \dot a a^{(3)} ) /a^4\cr
&+ \alpha_3 \hbar ( 3 {k^2}
 + 6 k {{\dot a}^2} + 3 {{\dot a}^4} ) / a^4
 + O(\hbar^2).\cr}
\eqn\Einsteintt$$
Here, $\rho_0$ is the radiation density at the time when the
scale factor is $a_0$.
The $a^{-4}$ dependence of the $\rho_0$ term follows directly
 from conservation of the stress-energy and the fact that the
the expectation value of the trace, $T_a{}^a$, vanishes to
lowest order.
The order $\hbar$ quantum corrections to this trace give
rise to the $\alpha_1$ and $\alpha_3$ terms. The values of
of the $\alpha$'s depend on which massless fields make up
the radiation.
In the spatially closed universe ($k = 1$), as a result of the
non-local Casimer vacuum energy, the value of $\rho_0$ will have
added to it a small constant value proportional to $\hbar$.$^{27,28}$
There is no additional non-local contribution to the stress-energy
because we are dealing with conformally invariant free radiation
fields in Robertson-Walker metrics, which are all conformally flat.
Before attempting to solve this equation, we reduce it to lower
order in time derivatives of $a(t)$,
so that all solutions are assured to be of physical significance.
\bigskip

\noindent {\bf{3.1\ \ Reduction of equation for scale factor}}

Multiplying Eq.\ \Einsteintt\  by $\hbar$ and working to first
order in $\hbar$ gives
$$\hbar \dot a^2 = - \hbar k  + \hbar {{\kappa \rho_0 a_0^4 }\over
        {3 {{a}^2}}} + \hbar {{\Lambda a^2}\over 3} + O(\hbar^2).
\eqn\aoneSquared$$
This expression can be used twice to find that
$$\eqalign{\hbar \dot a^4  &= \hbar {k^2} + \hbar {{2 {\Lambda}
\kappa  {\rho_0 {{{a_0}}^4}}}\over
        9} + \hbar {{ {{\kappa }^2} {{{\rho_0}}^2{{{a_0}}^8}}}\over
        {9 {{a}^4}}}\cr
&\qquad - \hbar {{2  k \kappa  {\rho_0 {{{a_0}}^4}}}\over
        {3 {{a}^2}}} - \hbar {{2 {\Lambda} k {{a}^2}}\over 3} +
      \hbar  {{{{{\Lambda}}^2} {{a}^4}}\over 9} + O(\hbar^2).\cr}
\eqn\aoneFour$$
Differentiating Eq.\ \aoneSquared\  leads to
$$\hbar \ddot a =
    - \hbar {{ \kappa  \rho_0 a_0{}^4 }\over
        {3 a^3}} + \hbar {\Lambda a\over 3} + O(\hbar^2).
\eqn\atwoRule$$
Here we have divided by $\dot a$, assuming that it is not zero in
the time interval of interest.
Making use of this result twice gives
$${{\hbar {{\ddot a}^2}} =
    {\hbar {{ {{\kappa}^2} {{{\rho_0}}^2}{{{a_0}}^8}}\over
        {9 {{a}^6}}} - \hbar {{2 {\Lambda}  \kappa
          {\rho_0 {{{a_0}}^4}}}\over {9 {{a}^2}}} + \hbar
      {{{{{\Lambda}}^2} {{a}^2}}\over 9}}} + O(\hbar^2).
\eqn\atwoSquared$$
We also need the third derivative, $\hbar a^{(3)}$, which is obtained
by differentiating Eq.\ \atwoRule :
$${{\hbar a^{(3)}} =
    {\hbar {{{\Lambda} \dot a}\over 3} + \hbar
      {{ \kappa  {\rho_0 {{{a_0}}^4}} \dot a}\over {{{a}^4}}}}}
       + O(\hbar^2).
\eqn\athreeRule$$

Using the above results repeatedly in Eq.\ \Einsteintt\  reduces
it to the first order differential equation,
$$\eqalign{0  =  &-\Lambda + {{3 k}\over {{{a}^2}}} +
   {{3 {{\dot a}^2}}\over {{{a}^2}}}
   - {{ \kappa  {\rho_0}{{{a_0}}^4}}\over {{{a}^4}}}
   - {\alpha_1} \hbar {{8 {\Lambda}  \kappa
             {\rho_0}{{{a_0}}^4}}\over {{{a}^4}}}\cr
&+ \alpha_3 \hbar \left( {{{{{\Lambda}}^2} }\over 3}
     + {{2 {\Lambda} \kappa {\rho_0}{{{a_0}}^4}}\over {3 {{a}^4}}}
     + {{ {{\kappa }^2} {{{\rho_0}}^2} {{{a_0}}^8}}\over {3 {{a}^8}}}
     \right)
     + O(\hbar^2).\cr}
\eqn\EinsteinttReduced$$
Except for the correction term involving $a^{-8}$, all of the
terms of order $\hbar$ can be absorbed by a renormalization
of the constants $\Lambda$ and $\kappa$. We make use of this
renormalization when we discuss below models with radiation,
spatial curvature and non-zero cosmological constant.
\bigskip
\vfil\eject

\noindent {\bf{3.2\ \ Spatially flat model with radiation and $\Lambda =0$}}

A simple but illuminating solution to the reduced
Friedmann-Lema\^\i tre-Einstein equation \releven\ is the spatially
flat $(k=0)$ case with zero cosmological
constant $(\Lambda =0)$ and pure radiation
$\left(\rho =\rho _0(a_0/ a)^4\right)$.
This should be a good approximation to our universe for one part of its
history, after any inflationary epochs which smoothed out inhomogeneities
but before massive fields cooled to non-relativistic temperatures.  The
classical solution, which has the scale factor grow as the square root of
cosmological time, begins with a curvature singularity, and expands
forever, becoming more and more flat.  Because the physical semiclassical
solutions are corrections to the classical solutions in powers of
curvature, we expect that at late times, when the classical solution is
nearly flat, the semiclassical corrections will be small.  At early times,
when the curvature is below the Planck scale but not above it, we expect
the semiclassical corrections to be significant.  At very early times,
however, when the classical curvature is near or above the Planck scale,
we expect that the semiclassical approximation will break down because
neglected higher order corrections would dominate.  We shall see how these
effects manifest themselves below.

For $\Lambda =0$, $k=0$, and $\rho =\rho _0(a_0/a)^4$, Eq.\
\releven\ reduces to
$$\dot a^2 = {\kappa\rho a^4_0\over{3a^2}} - \alpha_3\hbar
{\kappa^2\rho^2a_0^8\over{9a^6}} + O(\hbar^2) ,$$
which can be simplified by redefining the scale factor, cosmological time, and
correction constant $\a_3$, in dimensionless units
$${\tilde a}\equiv a\left( {4\kappa\rho_0a_0^4\over 3}\right)^{-{1\over 2}}
\qquad {\tilde t}\equiv t\left( {4\kappa\rho_0a_0^4\over 3}\right)^{-{1\over
2}} \qquad {\tilde \alpha _3}\equiv \a_3\left( {64\kappa\rho_0a_0^4\over
3}\right)^{-1} ,$$
giving
$${\dot {\tilde a}}^2 = {1\over{4{\tilde a}^2}} - \hbar{\tilde \alpha _3}
{1\over{{\tilde a}^6}} +O(\hbar^2), \eqn\a$$
where ${\dot {\tilde a}} = {d{\tilde a}\over{d{\tilde t}}}$.  This can be
solved iteratively with the ansatz
$${\tilde a}({\tilde t}) = {\tilde a}_0({\tilde t})+\hbar {\tilde
a}_1({\tilde t})+O(\hbar^2) . \eqn\b$$
Inserting Eq.\ \b\ into Eq.\ \a\ and expanding in powers of $\hbar$ gives
$$\eqalign{\hbar^0: &\quad {\dot {\tilde a}}^2_0 =
{1\over 4} {\tilde a}_0^{-2}\cr
\hbar^1: &\quad 2{\dot {\tilde a}}_0{\dot {\tilde a}}_1 =
- {1\over 2} {\tilde a}_1{\tilde a}_0^{-3}
+{\tilde \alpha _3} {\tilde a}_0^{-6}
.\cr} \eqn\c$$
The first equation gives the one parameter family of (classical) solutions
$${\tilde a}_0({\tilde t})=({\tilde t}-\tau_0)^{{1\over 2}} ,$$
where $\tau_0$ is a constant of integration.  When this is substituted
into the
second differential equation in \c , it gives the first order, linear,
inhomogeneous equation
$${\dot {\tilde a}}_1 + {1\over{2({\tilde t}-\tau_0)}} {\tilde a}_1 =
{\tilde \alpha _3} ({\tilde t}-\tau_0)^{-{5\over 2}} ,$$
which, for a given $\tau_0$, has a one parameter of family of solutions,
$${\tilde a}_1({\tilde t})=-{\tilde \alpha _3} ({\tilde
t}-\tau_0)^{-{3\over 2}} - {1\over 2} \tau_1 ({\tilde t}-\tau_0)^{-
{1\over 2}},$$
given by $\tau_1$, another constant of integration.  The full solution is
$${\tilde a}({\tilde t})=({\tilde t}-\tau_0)^{{1\over 2}} -\hbar{\tilde
\alpha _3} ({\tilde t}-\tau_0)^{-{3\over 2}} - {1\over
2} \hbar\tau_1({\tilde t}-\tau_0)^{-{1\over 2}} +O(\hbar^2) . \eqn\d$$
The apparently extra one parameter family of solutions (the freedom to specify
$\tau_1$ as well as $\tau_0$) is a result of the ambiguity in what it means to
specify a solution.  The freedom to specify both $\tau_0$ and $\tau_1$ is
equivalent to the freedom to specify one constant of integration, $\tau$, to
lowest and first order in $\hbar$.  This can be seen by expanding $\tau$ in
powers of $\hbar$, $\tau = \tau_0+\hbar\tau_1+O(\hbar^2)$, and using it to
write the solution \d\ as
$${\tilde a}({\tilde t})=({\tilde t}-\tau)^{{1\over 2}} -\hbar{\tilde \alpha
_3} ({\tilde t}-\tau )^{-{3\over 2}} +O(\hbar^2).$$
This freedom is related to the freedom of choosing different asymptotic
expansions for the solutions, and is discussed in more detail in Appendix B.

Another solution to Eq.\ \a , to the same order in $\hbar$, is
$$\tilde A({\tilde t})=[({\tilde t}-\tau )-2\hbar{\tilde \alpha _3}
({\tilde t}-\tau )^{-1}]^{{1\over 2}} + O(\hbar^2).$$
${\tilde a}$ and $\tilde A$ differ only by terms higher order in $\hbar$,
so either solution is as good as the other.  To the extent the two
solutions disagree, the semiclassical approximation cannot predict which
solution is more accurate.

Fig.\ 1 shows plots of ${\tilde a}$, $\tilde A$, and ${\tilde a}_0$
as functions of ${\tilde t}$, for two values of ${\tilde \alpha _3}$ (or,
equivalently, $\alpha_3$).  The three regimes referred to above can be
observed.  When the universe is nearly flat (for ${\tilde t}\; {\buildrel
>\over\sim}\; 1$), the semiclassical corrections are small.  For
intermediate scales
$(0.25\;{\buildrel <\over\sim}\; {\tilde t}\; {\buildrel <\over\sim}\;
0.75$, for the particular
${\tilde \alpha _3}$ plotted, ${\tilde \alpha _3} =\pm 0.01$), the
corrections are more substantial and cause noticeable deviations from the
classical solution.  For very early times $({\tilde t}\;{\buildrel
<\over\sim}\; 0.25)$, the corrections dominate the classical solution.
This, unfortunately, is the regime in which they cannot be trusted.  This
is made most obvious by noting the substantial differences (along the
${\tilde a}$ scale) between the solutions ${\tilde a}$ and $\tilde A$ for
small ${\tilde t}$.  Since they are both equally valid solutions to the
reduced semiclassical equations, to the extent they differ, the
predictions of either are not meaningful.  For positive (negative)
${\tilde \alpha _3}$ the effect of the corrections is to make the universe
larger (smaller) than it would have been at small times, but the most
dramatic predictions indicated by the solutions (and their plots) are in
regimes where the solutions should not be trusted.  All one can say is
that at small times and large curvature, semiclassical corrections are
important, and that at very small times and very large curvature, the
contribution from higher order corrections of quantum gravity are
necessary in order to make any meaningful predictions. The values chosen for
${\tilde \alpha _3}$ in Fig.\ 1 ($\pm 0.01$) have been chosen
extraordinarily large (though still well within the perturbative regime)
to demonstrate the qualitative effects of the quantum corrections. If one
were to choose $\alpha _3$ of the Planck scale and a standard cosmological
value for $\rho_0 a_0^4$, ${\tilde \alpha _3}$ might be as small as
$10^{-100}$.
\bigskip

\noindent {\bf{3.3\ \ Models with radiation, spatial curvature, and $\Lambda
\not= 0$}}

More general models with general spatial curvature  and cosmological
constant and radiation, are governed to first order in $\hbar$ by
Eq.\ \EinsteinttReduced . This can be rewritten in the form
$$\eqalign{0 \; &= \; { {3 \dot a^2}\over{a^2} } + {{3 k}\over{a^2} }
-\Lambda_r - { {\kappa_r  {\rho_0} {a_0}^4}\over {a^4} } +
  \alpha_3 \hbar { { {\kappa_r}^2 {\rho_0}^2 {a_0}^8}\over{3 a^8} }
   + O(\hbar^2),\cr} \eqn\EinsteinttRenormalized$$
where constant terms first order in $\hbar$ have been absorbed into
a renormalization of the gravitational and cosmological constants:
$$\Lambda_r \equiv \Lambda (1 - {1\over 3} \alpha_3 \hbar \Lambda),
\eqn\LambdaRenormalized$$
and
$$\kappa_r \equiv \kappa (1 - {2\over 3} \alpha_3 \hbar \Lambda_r
    + 8 \alpha_1 \hbar \Lambda_r).
\eqn\kappaRenormalized$$
With the exception of the final term, Eq.\ \EinsteinttRenormalized\ has the
same
form as in the corresponding classical Einstein equation, but with
$\kappa$ and $\Lambda$ having constant corrections of order $\hbar$.

Before numerically integrating Eq.\ \EinsteinttRenormalized ,
it is advantageous to carry out some analytic simplification.
Define the constants,
$$A \equiv {4\over 3} \Lambda_r,
\eqn\defA$$
$$B \equiv {4\over 3} \kappa_r \rho_0 {a_0}^4,
\eqn\defB$$
$$ C \equiv B A^{-1} - 4 k^2 A^{-2}.
\eqn\defC$$
Also define a dimensionless independent variable
$$ s \equiv A^{1/2} t,
\eqn\defs$$
and a function $g(s)$ by
$$ g \equiv a^2 - 2 k A^{-1}.
\eqn\defg$$
Then Eq.\ \EinsteinttRenormalized\  can be written as
$$({d\over ds} g)^2 - g^2 - C +
     \hbar \alpha_3 { {A^{-1} B^2} \over {4(g + 2 k A^{-1})^2} } = 0.
\eqn\EinsteinttSimplified$$
Notice that $C$ has the same value for $k = \pm 1$, so that the
solution of this equation without the final term of order $\hbar$
is the same for positive or negative spatial curvature, and is the
same as the classical solution with renormalized gravitational and
cosmological constants.

Probably the most interesting case to consider in more detail
is that of positive spatial curvature, $k=1$, with positive
cosmological constant, $\Lambda > 0$. This case includes both
the ``hesitation'' and ``turn-around'' models of Lema\^\i tre.

The hesitation model spends a long time near the configuration of
the Einstein static universe, during which density perturbations can
grow rapidly. Close relatives of this model, which do not require
a cosmological constant, have been recently considered in connection
with the growth of perturbations and galaxy formation.$^{24}$
Such models may be affected by quantum corrections in a manner
similar to that considered here.

The turn-around model with no singularity is of interest because
the perturbative approximation for the quantum correction is valid
during the entire evolution of the model. Over a long period of time
this quantum correction can cause a significant deviation of the
radius of the universe from that of the corresponding classical model.

The hesitation solution occurs when $C > 0$, and has zeroth order
solution of Eq.\ \EinsteinttSimplified ,
$$g = C^{1/2} \sinh(s)
\eqn\classicalHesitation$$
The turn-around solution occurs when $C < 0$, and has zeroth order
solution,
$$g = |C|^{1/2} \cosh(s)
\eqn\classicalTurnAround$$
In these solutions $s$ could be replaced by $s-s_0$, with $s_0$
constant. Also, since the zeroth order equation (for $g$) does not depend
on the sign of $k$, these classical solutions remain valid
when $k = -1$. However, the quantum correction does depend on the
sign of $k$; and we will only consider the case of positive
spatial curvature.

We next consider the corrections to the turn-around model,
and then the corrections to the hesitation model.
\bigskip

{\bf{3.3.1\ \ Turn-around or bounce universe}}

Dividing Eq.\ \EinsteinttSimplified\  by $|C|$, and defining
the dimensionless function
$$G(s) = |C|^{-1/2} g(s),
\eqn\defG$$
and the dimensionless constants
$$u = A^{-1}|C|^{-1/2},
\eqn\defu$$
which regulates the abundance of radiation relative to the cosmological
constant, and
$$v = (1/4) \hbar \alpha_3 B^2 / (A C^2),
\eqn\defv$$
which regulates the quantum corrections,
we obtain in the case of the turn-around solution ($C<0$):
$$  ({d\over ds} G(s) )^2 - G(s)^2 + 1 +
     v (G(s) + 2 u)^{-2} = 0.
\eqn\turnAround$$
(The hesitation model obeys the same equation with $+1$ replaced
by $-1$.) The initial conditions for a turn-around solution
are that $G > 1$ and ${d\over ds} G < 0$, so that $G$ will approach
a minimum value near $1$ and then increase.
The perturbative solution will remain valid at all times for which
$$v (G(s) + 2 u)^{-2} \ll 1.
\eqn\validity$$

The relationship between $a$ and $G$ can be written as
$$(A^{1/2} a)^2 = (u^{-1} G + 2)^{1/2}.$$
The upper curves in Fig.~2 show the radius $a$ as
a function of time $t$, with both $a$ and $t$ measured in units of
of $A^{-1/2}=(4\Lambda_r/3)^{-1/2}$. Both curves have $u=0.5$.
The lower curve has $v=0$.
The upper curve includes the quantum corrections
coming from $\alpha_3$, and has $v=0.4$. The boundary condition
for both curves was that $G(1) = \cosh(1)$.
Again, as in Fig.\ 1, the numerical value of the quantum correction,
in this case $v=0.4$, has been chosen to be much larger than one would
expect from Planck scale contributions, in order to emphasize the
qualitative effects of the corrections.

The curves were obtained by numerical
integration of the equation for ${d\over ds} G(s)$.
In solving Eq.~\turnAround\ for ${d\over ds} G(s)$, the positive
square-root was used in the range form $s=1$ to $s=0.09233$
at which $G(s)$ has its minimum value of $1.0467$. The negative
square-root was used for smaller values of $s$, since $G(s)$ is
a decreasing function of $s$ in that range. The correction term
in Eq.~\turnAround\ remains smaller than $0.1$ and has its largest
value when $G$ goes through its minimum. Thus, the perturbative
solution is valid for all values of $s$.

The lower curve is the same as the
classical solution with $\Lambda = \Lambda_r$, $\kappa=\kappa_r$.
It corresponds to the solution given in Eq.~\classicalTurnAround ,
and coincides with the quantum corrected solution at $t=1$ (in units of
$A^{-1/2}$).  It is clear that as a result of the deviation introduced
near the minimum radius by the the quantum correction, the two models have
increasingly different radii as one goes back into the past.

\bigskip

{\bf{3.3.2\ \ Semiclassical Hesitation Universe}}

 From the previous examples, one can see that it is difficult to find
examples of semiclassical corrections that are not overwhelmingly small.
The last example used the exponential growth of the scale factor to
magnify small correction terms.  One instance semiclassical corrections
might have a macroscopic impact is when the classical field equations
describe an unstable system.  Even initially small quantum effects could
dominate the behavior at late times.

One important example of an unstable cosmological solution to the classical
Einstein equations is the unstable Einstein static universe.  Historically,
Einstein looked for a static cosmological solution to describe an eternal
universe, leading him to modify his equations with a cosmological constant.
This
solution is unstable, however, and small changes in initial data cause the
universe to expand forever at an exponential rate (due to the cosmological
constant) or to collapse to a curvature singularity.  One would expect quantum
corrections to have similar, drastic effects.

The simplest Einstein static universe is spatially closed $(k=1)$, contains a
radiation fluid, with density $\rho =\rho _0(a_0/a)^4$, and cosmological
constant, $\Lambda$, both carefully chosen such that $\Lambda
=9/(4\kappa\rho_0a_0^4)$.  Defining a natural length scale, $L_0$, such that
$L_0
= {\sqrt{3/\Lambda}} = {\sqrt{4\kappa\rho_0a_0^4/3}}$, we can write the
classical
Einstein equation as
$$\eqalign{O(\hbar ) &= \dot a^2 - {\Lambda\over 3} a^2 + k -
{\kappa\rho_0a_0^4\over{3a^2}}\cr
&= \dot a^2 - L_0^{-2}a^2+1-{1\over{4L_0^{-2}a^2}}\cr} . \eqn\hesone$$
This has the solutions
$$a^2 = \left\{ \matrix{ {L^2_0/2}\cr\m
{(L^2_0/2)} \left[ 1-\exp\left( \mp {2(t-t_0)/{L_0}}\right)\right]\cr\m
{(L^2_0/2)} \left[ 1+\exp\left( \pm {2(t-
t'_0)/{L_0}}\right)\right]\cr}\right\} +O(\hbar ) , \eqn\hestwo$$
plotted in Fig.\ 3.  The first solution is the Einstein static universe.  The
second begins at a singularity and asymptotically approaches the Einstein
static
universe at late times (the positive exponential is the time-reversed
solution).
The third spends an infinite time near the Einstein static universe, but pulls
away and ends in an infinite inflationary epoch (the negative exponential is
the
time-reversed solution).  The first and second solutions (and the time-reversed
third) are unstable.

The behavior of the semiclassical corrections should reflect the instability of
the classical solutions.  The semiclassical Einstein equation of Eq.\ \releven\
becomes
$$O(\hbar )=\dot a^2 - L_0^{-2}a^2+1-{1\over{4L_0^{-2}a^2}}-
6\alpha_1\hbar{1\over{a^2}}+\alpha_3\hbar {1\over{a^2}} \left( L_0^{-2}a^2 +
{1\over{4L_0^{-2}a^2}}\right)^2 . \eqno{(3.34a)}$$
This can be solved iteratively, as above, by expanding the solutions as a
series
in $\hbar$:
$$a=a_{cl}+\hbar a_1+O(\hbar^2) , \eqno{(3.34b)}$$
where $a_{cl}$ is one of the classical solutions in Eq.\ \hestwo .  Only the
second and third solutions of Eq.\ \hestwo\ generate solutions to the
semiclassical equations in general.  [It should not be surprising that the
constant solution has no semiclassical counterpart.  The effect of the
semiclassical corrections in Eq.\ (3.34a), if viewed as a Newtonian energy
equation, is to change the potential term without changing the total energy.
The
energy corresponding to a static configuration without the corrections will
not,
in general, correspond to a static configuration when the potential is changed;
to regain a static solution one must change the initial matter density
appropriately.]  The equation of motion for $a_1$, obtained by inserting Eq.\
(3.34b) into Eq.\ (3.34a), is a linear, first order equation, and can be solved
in
closed form.

The semiclassical solution generated by the initially singular classical
solution
is
$$\eqalign{a^2 &= {L^2_0\over 2} (1-\exp [-2(t-\tau )L_0^{-1}])\cr
&\quad \times \biggl[ 1+3\alpha_1\hbar L_0^{-2} {\exp [2(t-\tau )L_0^{-
1}]\over{1-\exp [-2(t-\tau )L_0^{-1}]}}\cr
&\qquad - {\alpha_3\over 4} \hbar L_0^{-2} \biggl( 2 {\exp [2(t-\tau )L_0^{-
1}]\over{1-\exp [-2(t-\tau )L_0^{-1}]}} - {\exp[-2(t-\tau
)L_0^{-1}]\over{(1-\exp
[-2(t-\tau )L_0^{-1}])^2}}\cr
&\qquad\qquad\qquad\quad +8(t-\tau )L_0^{-1} {\exp[-2(t-\tau
)L_0^{-1}]\over{1-\exp [-
2(t-\tau )L_0^{-1}]}}\cr
&\qquad\qquad\qquad\quad +3 {\exp [-2(t-\tau )L_0^{-1}]\over{1-\exp [-2(t-\tau
)L_0^{-
1}]}} \ln (1-\exp [-2(t-\tau )L_0^{-1}])\biggr)\biggr]\cr
&\qquad\qquad\qquad\qquad\qquad\qquad\qquad\qquad\qquad\qquad\qquad +O(\hbar^2)
,\cr} \eqn\hesthree$$
where $\tau = t_0+\hbar t_1+O(\hbar^2)$.  As shown in Appendix A, the form of
the solution \hesthree\ is ambiguous up to the addition of $O(\hbar )$ terms
proportional to $\dot a_{cl}$, arising from shifts $O(\hbar )$ in the initial
time $(t_0\rightarrow\tau =t_0+\hbar t_1)$.  One may always choose a $t_1$ such
that the coefficient of the ambiguous term is zero.
This solution is plotted in Fig.\ 4 for $\alpha_1\hbar L_0^{-2} = 0.0001$ and
$\alpha_3\hbar L_0^{-2} = -0.0001$ (though the qualitative behavior is
independent
of those values, as long as $6\alpha_1-\alpha_3 >0$).
As in the case of the spatially
flat, radiation-filled universe above, the semiclassical approximation breaks
down when too close to the initial singularity.  At later times, where the
semiclassical approximation is good, the effect of
the corrections is to pull away from the Einstein static solution and begin an
inflationary epoch.  At very late times, when the correction terms dominate the
classical solution completely, the corrections are untrustworthy, but at these
late
times the other (late-time-de Sitter) semiclassical solution exhibits the same
inflationary behavior, and in a trustworthy regime. One can hope to match the
two
semiclassical solutions in the intermediate regime, where both are valid.  We
do this below.

The semiclassical counterpart to the late-time-de Sitter classical solutions of
Eq.\ \hestwo\ takes the form
$$\eqalign{a^2 &= {L^2_0\over 2} (1+\exp [-2(t-\tau ')L_0^{-1}])\cr
&\quad \times \biggl[ 1-3\alpha_1\hbar L_0^{-2} {\exp [2(t-\tau ')L_0^{-
1}]\over{1+\exp [-2(t-\tau ')L_0^{-1}]}}\cr
&\qquad + {\alpha_3\over 4} \hbar L_0^{-2} \biggl( 2 {\exp [2(t-\tau ')L_0^{-
1}]\over{1+\exp [-2(t-\tau ')L_0^{-1}]}} - {\exp[-2(t-\tau
')L_0^{-1}]\over{(1+\exp
[-2(t-\tau ')L_0^{-1}])^2}}\cr
&\qquad\qquad\qquad\quad -8(t-\tau ')L_0^{-1} {\exp[2(t-\tau
')L_0^{-1}]\over{1+\exp [
2(t-\tau ')L_0^{-1}]}}\cr
&\qquad\qquad\qquad\quad +3 {\exp [2(t-\tau ')L_0^{-1}]\over{1+\exp [2(t-\tau
')L_0^{-
1}]}} \ln (1+\exp [-2(t-\tau ')L_0^{-1}])\biggr)\biggr]\cr
&\qquad\qquad\qquad\qquad\qquad\qquad\qquad\qquad\qquad\qquad\qquad +O(\hbar^2)
,\cr} \eqn\hesfour$$
where $\tau ' = t'_0+\hbar t'_1+O(\hbar^2)$ and $t'_1$ is chosen analogously to
$t_1$ above.  Eq.\ \hesfour\ is plotted in Fig.\ 4 for two values of $\alpha_1$
and
$\alpha_3$.  At late times, the corrections are very small compared to the
classical
solution.  At intermediate times, the corrections are small but non negligible,
and at early times the corrections are so large as to be untrustworthy $(\hbar
a_1/a_{cl}\; {\buildrel >\over\sim}\; 1$).

Because the semiclassical solutions of Eq.\ \hesthree\ and Eq.\ \hesfour\ are
valid in different regimes, it is important to ask if there is any overlap of
the
regimes where both solutions are valid.  Furthermore, if there is such a
regime,
perhaps the solutions can be smoothly joined, corresponding to a universe
beginning at large curvature near a singularity, flattening off at nearly
constant scale factor for an extended period of time, and then proceeding to
inflate in a de Sitter-like phase.  This would correspond to a classical
``hesitation'' universe in which the matter density (or cosmological constant)
is slightly greater than necessary for the Einstein static universe.

For $6\alpha_1-\alpha_3 >0$ (the parameter range for which the solution of Eq.\
\hesthree\ is expanding at late times) there is an overlap region in which we
can
match the solution of Eq.\ \hesthree\ to the solution of Eq.\ \hesfour , as
shown
in Fig.\ 4 by using the freedom to set the base times $(\tau$ and $\tau '$) of
each solution individually.  The matching can always be done smoothly, since
the
curves of $a$ cross for all values of $\tau -\tau '$, and we can adjust $\tau -
\tau '$ such that $\dot a$ is continuous (sufficient for
matching solutions of a first order equation).

Furthermore, $\ddot a$ is discontinuous only by terms $O(\hbar^2)$.  The
matching
can be done in regions where $\hbar a_1/a_{cl} \ll 1$ for both solutions for a
wide variety of parameters (such that $6\alpha_1-\alpha_3>0$).  We may
naturally
interpret this joining of matched solutions as a unique solution to the
semiclassical equation that is everywhere perturbatively valid (except the
region
near the initial singularity).  The time of hesitation, $t_h\approx\tau
'-\tau$,
determined by the matching conditions, is logarithmically related to the
coefficients of the semiclassical corrections:  $t_h\approx L_0\ln\left[
(6\alpha_1-
\alpha_3)\hbar L_0^{-4}\right]$, due to the exponential time dependence of the
semiclassical solutions.

The only potential obstacle to this interpretation is that, although ${\hbar
a_1/{a_{cl}}} \ll 1$ where the joining is done, ${\hbar\dot a_1/{\dot
a_{cl}}} \approx 1$.  We feel that this is no reason to doubt the validity of
the
joining, however, since ${\hbar\dot a_1/{\dot a_{cl}}}$ becomes large due to
$\dot a_{cl}$ vanishing, not due to $\hbar\dot a_1$ becoming large.
\vfil\eject

\chapternumber=4
\makechapterlabel
\equanumber=0

\centerline{\bigbf{4.  Conformally Flat Corrections}}

A conformally flat background spacetime (for which Eq.\Einstein\ gives the
general
form of the local corrections to the stress energy) has a metric tensor related
to the Minkowski metric tensor by
$$g^{cl}_{ab}=e^{2S}\eta_{ab} , \eqn\gcone$$
where $e^{2S}$ is the conformal factor (a general scalar function), and
$\eta_{ab}$ is the Minkowski tensor (note that the conformal transformation is
not, in general, a diffeomorphism).  Conformally flat metrics are special in
this
sense --- the entire metric (a symmetric tensor with 6 independent components)
is completely specified by a single scalar function on the spacetime.

As Eq.\ \gcone\ is written, it is a tensor equation on the spacetime with
classical metric.  Here the Minkowski tensor $\eta_{ab}$ is a tensor function
defined on the physical spacetime.  Since $\eta_{ab}$ is also the metric of
flat
Minkowski spacetime (or a piece of Minkowski spacetime), Eq.\ \gcone\ implies
that there is a map from the physical spacetime with metric $g^{cl}_{ab}$ to an
unphysical flat spacetime with metric $\eta_{ab}$, and the conformal factor
$e^{2S}$ can be viewed as a scalar function on either spacetime.  If the
conformal factor is known on the unphysical flat spacetime, this knowledge can
be exploited to simplify the calculation of the semiclassical corrections,
i.e.\
to do all the calculations on the unphysical flat background.

We expand the general semiclassical Einstein equations for a conformally flat
background
$$G_{ab}(g_{cd}) + \alpha_1\hbar\; ^{(1)}H_{ab}(g_{cd})+\alpha_3\hbar\;
^{(3)}H_{ab}(g_{cd})-\kappa T_{ab}(g_{cd})=O(\hbar^2), \eqn\gctwo$$
where
$$g_{cd}=g^{cl}_{cd}+\hbar h_{cd}+O(\hbar^2) .\eqn\gcthree$$
This has the classical lowest order expansion
$$G_{ab}(g^{cl}_{ab})-\kappa T_{ab}(g^{cl}_{ab})=O(\hbar ), \eqn\gcfour$$
which is already known if $S$ is known.  The equation first order in
$\hbar$ is
$$\hbar {\delta G_{ab}\over{\delta g_{ef}}} (g^{cl}_{cd})h_{ef}+\alpha_1\hbar\;
^{(1)}H_{ab}(g^{cl}_{cd})+\alpha_3\hbar\; ^{(3)}H_{ab}(g^{cl}_{cd})
-\hbar\kappa
{\delta T_{ab}\over{\delta g_{ef}}} (g^{cl}_{cd})h_{ef}=O(\hbar^2)
\eqn\gcfive$$
where
$$\eqalign{ {\delta G_{ab}\over{\delta g_{ef}}} h_{ef} \equiv &- {1\over 2}
\nabla_a\nabla_bh-{1\over 2} \square h_{ab}+\nabla_c\nabla_{(a}h_{b)}{}^c -
{1\over 2} R^{cl}h_{ab}\cr
&- {1\over 2} g^{cl}_{ab} \left[ -\square
h+\nabla_c\nabla_dh^{cd}+R^{cl}_{cd}h^{cd}\right]\cr} \eqn\gcsix$$
and all derivatives and raising of indices are with respect to the classical,
physical metric.  We have not explicitly expanded the
last term of Eq.\ \gcfive, the functional derivative of
the stress-energy tensor, since its functional dependence on the metric depends
on the particular form of matter present.  If the functional dependence
is known (as is often the case) then it is straightforward
to calculate.

This is a set of second order, linear, inhomogeneous equations for $h_{ab}$.
The
second order equation produces a two-parameter family of solutions (just as for
the classical equation).  The freedom to choose two additional free parameters
in the semiclassical solutions arises from the freedom to specify the two
parameters of the solution to the full semiclassical equations at both
classical
and semiclassical order independently.

The power of Eq.\ \gcone\ is most apparent when the conformal factor $e^{2S}$
is known as a function of the fictitious flat spacetime.  All semiclassical
calculations can be performed on the flat spacetime instead of the physical
spacetime.

Then Eq.\ \gcsix\ becomes
$$\eqalign{ {\delta G_{ab}\over{\delta g_{ef}}} h_{ef} &= {1\over 2} e^{-2S}
[2h_{c(a;b)}{}^c-h_{;ab}-h_{ab;c}{}^c+8S_{;c}{}^ch_{ab}+4S_{;c(a}h_{b)}{}^c\cr
&\qquad - 4S_{;(a}h_{b)c;}{}^c+4S_{;(a}h_{;b)}+2S_{;c}h_{ab;}{}^c\cr
&\qquad +2S_{;a}S_{;b}h+S_{;c}S_{;}{}^ch_{ab}-8S_{;c}S_{;(a}h_{b)}{}^c\cr
&\qquad +\eta_{ab}(-h_{cd;}{}^{cd}+h_{;c}{}^c+2S_{;c}h_;{}^c +2S_{;c}S_;{}^ch)]
,\cr} \eqn\gcseven$$
and the inhomogeneous terms are
$$\eqalign{^{(1)}H_{ab}(g^{cl}_{cd}) &=\; ^{(1)}H_{ab}(e^{2S}\eta_{cd})\cr
&= 6e^{-2S}\times [-12S_{;c}S_{;a}S_{;b}S_;^c-
12S_{;a}S_{;b}S_{;c}{}^c+12S_{;c}S_{;b}S_{;a}{}^c\cr
&\qquad +12S_{;c}S_{;a}S_{;b}{}^c-
4S_{;ca}S_{;b}{}^c+6S_{;b}S_{;c}{}^c{}_a+6S_{;a}S_{;c}{}^c{}_b\cr
&\qquad -4S_{;c}S_{;a}{}^c{}_b-
2S_{;c}{}^c{}_{ab}+\eta_{ab}(3S_{;c}S_{;d}S_:^cS_;^d-3S_{;c}{}^cS_{;d}{}^d\cr
&\qquad -12S_{;c}s_{;d}S_;^{cd}+4S_{;cd}S_;^{cd}-
2S_{;c}S_;^c{}_d{}^d+2S_{;c}{}^c{}_d{}^d)]\cr} \eqn\gceight$$
$$\eqalign{^{(3)}H_{ab}(g^{cl}_{cd}) &=\; ^{(3)}H_{ab}(e^{2S}\eta_{cd})\cr
&= e^{-2S}[-4S_{;c}s_{;a}S_{;b}S_;{}^c-
4S_{;a}S_{;b}S_{;c}{}^c+4S_{;c}{}^cS_{;ab}\cr
&\qquad \quad\; +4S_{;c}S_{;b}S_{;a}{}^c+4S_{;c}S_{;a}S_{;b}{}^c-
4S_{;ca}S_{;b}{}^c\cr
&\qquad \quad\; +\eta_{ab}(S_{;c}S_{;d}S_;{}^cS_;{}^d-2S_{;c}{}^cS_{;d}{}^d-
4S_{;c}S_{;d}S_;{}^{cd}+2S_{;cd}S_;{}^{cd})].\cr} \eqn\gcnine$$
where semicolons refer to derivatives covariant with respect to the unphysical
flat metric and all raising and lowering of indices in Eqs.\ \gcseven
-\gcnine\ are with $\eta_{ab}$.  For a given stress energy, it is
straightforward
to put the last term of Eq.\ \gcsix\ into a similar form.  By putting
Eq.\ \gcsix\ in this form, the problem has now been simplified from solving
a second order partial differential equation in curved spacetime to solving a
second order partial differential equation in flat spacetime.

Useful formulas analogous to Eqs.\ \gceight -\gcnine , but where the covariant
derivatives and raising and lowering of indices are performed with the physical
background metric have also been calculated.$^{19}$  These would be helpful in
the case that the map from the physical spacetime to the unphysical spacetime
implied by \gcone\ were not known explicitly.
\vfil\eject

\chapternumber=5
\makechapterlabel
\equanumber=0

\centerline{\bigbf{5.  General Case}}

The previous examples involved Friedmann-Robertson-Walker,
Lema\^\i tre, and other conformally flat spacetimes.
We now turn to general spacetimes. The perturbative constraints in the
general case can be used to express the curvature tensors appearing
in the first-order quantum corrections (i.e., the $H$'s)
in terms of the lowest-order (classical) stress-energy tensor
$T_{ab}$ of the matter. Usually this lowest-order $T_{ab}$ involves
fewer derivatives of the metric than do the curvature tensors, so that
this procedure results in an equation with fewer derivatives than
the original. For example, for a classical fluid or a minimally coupled
scalar field, the resulting equation contains no more than second
derivatives of the metric.

The reduction process must be modified
when the stress-energy tensor contains explicit curvature terms, as for
a conformally coupled scalar field. One way to deal with such a case is
to evaluate the curvature tensor appearing in $T_{ab}$ using the lowest
order classical solution, since only that will contribute to the
correction terms of order $\hbar$.

For simplicity, we will suppose that a
cosmological constant term, if present in the Einstein equation,
is included in the definition of $T_{ab}$. Such a term in $T_{ab}$
involves no derivatives of the metric.

The perturbatively constrained equations should
not have the instabilities exhibited when one
tries to integrate numerically higher derivative equations.
Such instabilities can be produced by the tendency for the
growing or runaway solutions in the enlarged solution space to dominate
the nearly classical perturbative part of solutions. The runaway
solutions will not be present in the solution space after the reduction
to lower derivatives.

The first-order perturbative constraint coming from the Einstein
equations with quantum corrections, Eq.\ \one, is
$$  \hbar G_{ab} = \hbar \kappa T_{ab} + O(\hbar^2),
\eqn\fiveone$$
where $T_{ab}$ is the zero-order contribution of the stress-energy tensor.
It follows that
$$  \hbar R = - \hbar \kappa T + O(\hbar^2),
\eqn\pertone$$
where $T = T_a{}^a$.
Also,
$$  \hbar R_{ab} = \hbar \kappa (T_{ab} - (1/2) g_{ab} T) + O(\hbar^2).
\eqn\perttwo$$
Substituting these perturbative constraints into the
expressions for the ${}^{(i)} H_{ab}$ gives (in four dimensions)
the following results.
$$  {}^{(1)} H_{ab} = \kappa \left(-2 T_{;ab}
     - (1/2) g_{ab} \kappa T^2 + 2 g_{ab} T_{;p}{}^p +
     2 \kappa T T_{ab}\right) + O(\hbar),
\eqn\HoneT$$
$$\eqalign{ {}^{(2)} H_{ab} &= \kappa\,\bigl(-T_{;ab}
     + T_{ab;p}{}^p + 2 T_a{}^p{}_{;bp} - (1/2) \kappa g_{ab} T^2\cr
&\qquad + 2 \kappa T T_{ab} - 2 \kappa T_{pb} T_a{}^p
     + (1/2) \kappa g_{ab} T_{pq} T^{pq}\bigr) + O(\hbar),\cr}
\eqn\HtwoT$$
and
$$\eqalign{ {}^{(3)} H_{ab} &= \kappa\,\bigl(-(1/6) g_{ab} \kappa T^2
      + (1/3) \kappa T T_{ab}\cr
&\qquad - \kappa T_{pb} T_a{}^p
      + (1/2) \kappa g_{ab} T_{pq} T^{pq}\bigr) + O(\hbar).\cr}
\eqn\HthreeT$$
Then the perturbatively constrained Einstein equation with
quantum corrections in a general spacetime is given by
Eqs.\ \rtwo -\rsix\ with these values for ${}^{(1)} H_{ab}$
and ${}^{(2)} H_{ab}$ (with $\Lambda = 0$).
Only the
lowest order or classical $T_{ab}$ appears in these quantum
correction terms, and if there is a classical cosmological
constant present, then it is included in the definition of the
lowest order $T_{ab}$.
In general, the stress tensor contribution
on the right-hand-side of the semiclassical Einstein equation
will include non-local state-dependent contributions,
such as those coming from gravitationally-induced particle creation
and other effects. Local state-independent quantum corrections
to the stress-energy tensor are, of course,
already included with the local correction
terms on the left-hand-side. If the zeroth order stress-tensor
has vanishing trace $T$, as for radiation or massless particles,
then the equations simplify considerably.

In a conformally flat spacetime, ${}^{(2)} H_{ab}$ is replaced
by ${}^{(3)} H_{ab}$, and further simplification may occur
if the field is a massless conformally invariant field. The way in which
this occurs was already discussed in the Introductory section.

Similar expressions for ${}^{(1)} H_{ab}$ and ${}^{(2)} H_{ab}$
were also obtained by Bel and Sirousse-Zia.$^7$
\bigskip

\noindent{\bf{5.1\ \ Spherical Body With Quantum Corrections}}

We next make use of the previous expressions for the correction terms, to
write the equations governing the quantum corrections to the gravitational
field of a static spherical body. These equations are in a form suitable
for numerical integration..

We first calculate the local state-independent quantum corrections which
enter into the Einstein equations for the most general spherically
symmetric spacetime, which has the line element
$$ds^2 = -B(r) dt^2 + A(r) dr^2 + r^2 d\theta^2 +
                           r^2 \sin^2(\theta) d\phi^2,
\eqn\SphericalMetric$$
Here $A$ and $B$ will consist of a classical part and a quantum
correction of order $\hbar$.

We will assume that the matter is described by a perfect fluid
energy-momentum tensor,
$$T^{ab} = p g^{ab} + (p + \rho) u^a u^b,
\eqn\FluidStressT$$
where $u^a$ is the four-velocity $dx^a/d\tau$ of the fluid volume
element, and the proper pressure $p$ and $\rho$ are functions only of
the radial coordinate $r$. Since the fluid is static, we have
$u^r = u^\theta = u^\phi = 0$, and $u^t = B(r)^{-1/2}$.

 From the previous section, we have the expressions for
${}^{(1)} H_{ab}$, ${}^{(2)} H_{ab}$, and
${}^{(3)} H_{ab}$ in terms of the classical fluid $T_{ab}$.
Here we give the result of the calculation of those
expressions (this is a lengthy calculation, but is simpler than calculating
the quadratic curvature tensor expressions directly; in addition, because
the perturbative constraints have been used, no higher than second
derivatives appear in the result).

One finds for the first correction term,
$$\eqalign{ {}^{(1)} H_{rr} &=  \kappa\,\bigl[-(1/2) \kappa \rho^2 A
     + \kappa \rho A p + (3/2) \kappa A p^2\cr
&\quad + 12 p'/r + 3 p' B'/B\cr
&\quad - 4 \rho'/r - 3 \rho' B'/B \bigr]
     + O(\hbar),\cr}
\eqn\Honeprhooneone$$
$$\eqalign{ {}^{(1)} H_{\theta \theta} &=  \kappa\,\bigl[-(1/2) \kappa r^2
\rho^2
     + \kappa r^2 \rho p + (3/2) \kappa r^2 p^2\cr
&\quad + 6 r p'/A - 3 r^2 p' A'/A^2\cr
&\quad + 3 r^2 p' B'/(AB) + 6 r^2 p''/A\cr
&\quad - r^2 \rho' B'/(AB) - 2 r^2 \rho''/A\cr
&\quad - 2 r \rho'/A + r^2 \rho' A'/A^2 \bigr]
     + O(\hbar),\cr}
\eqn\Honeprhotwotwo$$
$$  {}^{(1)} H_{\phi \phi}  = \sin^2\theta\ \,{}^{(1)} H_{\theta \theta}
\qquad \qquad \qquad ,
\eqn\Honeprhothrthr$$
and
$$\eqalign{ {}^{(1)} H_{tt} &= \left[3 \kappa B/(2 r A^2)\right]
        \bigl[-\kappa r \rho^2 A^2 + 2\kappa r \rho p A^2\cr
&\quad +3 \kappa r p^2 A^2 - 8 p' A\cr
&\quad + 2 r p' A' - 4 r p'' A\cr
&\quad + 4 \rho'A - r\rho'A'\cr
&\quad + 2 r\rho''A \bigr] + O(\hbar).\cr}
\eqn\Honeprhofourfour$$

The second correction term is found to be the following:
$$\eqalign{  {}^{(2)} H_{rr} &=  \kappa\,\bigl[ \kappa p^2 A
     + \kappa \rho A p - p B'^2/B^2\cr
&\quad - \rho B'^2/B^2 + 2 p'/r + (3/2) p' B'/B\cr
&\quad - (1/2) \rho' A'/A + \rho' B'/B
                              + \rho'' \bigr] + O(\hbar),\cr}
\eqn\Htwoprhooneone$$
$$\eqalign{ {}^{(2)} H_{\theta \theta} &=  \kappa\,\bigl[ \kappa r^2 p^2
      + \kappa r^2 \rho p + r p'/A\cr
&\quad - (1/2) r^2 p' A'/A^2
      + (1/2) r^2 p' B'/(AB)\cr
&\quad + r \rho'/A + r^2 p''/A \bigr] + O(\hbar),\cr}
\eqn\Htwoprhotwotwo$$
$$  {}^{(2)} H_{\phi \phi}  = \sin^2\theta\ \,{}^{(2)} H_{\theta \theta}
\qquad \qquad \qquad ,
\eqn\Htwoprhothrthr$$
and
$$\eqalign{ {}^{(2)} H_{tt} &=  \kappa\,\bigl[ 3 \kappa B p^2
                  + 3 \kappa B p \rho - 2 p B'/(r A )\cr
&\quad - 2 \rho B'/(r A ) + (1/2)p A' B'/A^2\cr
&\quad + (1/2)\rho A' B'/A^2 - (1/2) p B'^2/(AB)\cr
&\quad - (1/2) \rho B'^2/(AB) + (1/2) p' B'/A\cr
&\quad + 2 \rho' B/(rA) - (1/2) \rho' B A'/A^2\cr
&\quad - p B''/A - \rho B''/A\cr
&\quad + \rho'' B/A \bigr] + O(\hbar).\cr}
\eqn\Htwoprhofourfour$$

For the third correction term, which appears in place of
${}^{(2)} H_{ab}$ in a conformally flat spacetime,
the final result is much simpler:
$$ {}^{(3)} H_{rr}  =
        (1/3) \kappa^2 \rho (\rho + 2 p ) A + O(\hbar),
\eqn\Hthrprhooneone$$
$$  {}^{(3)} H_{\theta \theta}  = (r^2/A)\ {}^{(3)} H_{rr} + O(\hbar),
\eqn\Hthrprhotwotwo$$
$$  {}^{(3)} H_{\phi \phi}  = \sin^2\theta\ \,{}^{(3)} H_{\theta \theta}
\qquad \qquad \qquad ,
\eqn\Hthrprhothrthr$$
and
$$  {}^{(3)} H_{tt}  =
        (1/3) \kappa^2 \rho^2 B + O(\hbar),
\eqn\Hthrprhofourfour$$

It is interesting that ${}^{(3)} H_{ab}$ is in the form of a
perfect fluid energy-momentum tensor. It can be absorbed into the change,
$T_{ab} \rightarrow T_{ab} + \Delta T_{ab}$, with
$$\Delta T_{ab} = \Delta p g_{ab} +
                       (\Delta p + \Delta \rho) u_a u_b + O(\hbar^2),
\eqn\Tabthree$$
where $u_a$ is as before,
$$\Delta p = -(1/3) \alpha_3 \hbar \kappa \rho (\rho + 2 p),
\eqn\pthree$$
and
$$\Delta \rho = -(1/3) \alpha_3 \hbar \kappa \rho^2.
\eqn\rhothree$$
Therefore, in the case of a conformally flat metric,
the $\alpha_3 \hbar {}^{(3)} H_{ab}$ quantum correction term
in the Einstein equation, can be absorbed into a redefinition
of the pressure and density:
$$p \rightarrow p - (1/3) \alpha_3 \hbar \kappa \rho (\rho + 2 p)
                    + O(\hbar^2),
\eqn\pnew$$
and
$$\rho \rightarrow \rho - (1/3) \alpha_3 \hbar \kappa \rho^2 + O(\hbar^2).
\eqn\rhonew$$

The semiclassical Einstein equations with quantum corrections in
general have the form,
$$\eqalign{R_{ab} &- (1/2) g_{ab} R\cr
&+ \alpha_1\ \hbar\, {}^{(1)} H_{ab} +
         \alpha_2\ \hbar\, {}^{(2)} H_{ab} +
     \alpha_3\ \hbar\, {}^{(3)} H_{ab} + O(\hbar^2) = \kappa {\cal T}_{ab},\cr}
\eqn\Einsteintwo$$
where ${\cal T}_{ab}$ includes classical matter contributions and
the lowest order state-dependent part of the expectation value of
quantum matter fields.  The
state-independent local quantum corrections of order $\hbar$
are included in the $H$ terms on the left-hand-side. It is understood
that we may set $\alpha_3 = 0$, except when the metric is conformally flat. In
the latter case, it is understood that $\alpha_2 = 0$, since the first
two corrections are then proportional to one another. In the conformally
flat case, the $\alpha_3$ term arises from the state-dependent part
of the quantum stress-energy.

With the expressions given above for the $H$'s, the Einstein equations
are now easily written down for the general spherically symmetric metric.
Only the state-dependent part of the expectation value of the
quantum stress-energy tensor requires further work to calculate, but this
will not increase the order of the highest metric derivative in most
cases, so that the perturbative constraints have succeeded in reducing
the semiclassical Einstein equations with quantum corrections to
second order equations having the standard initial data. These equations
are thus in suitable form for numerical integration. We will not carry
that out here, but plan to return to it in a later paper.
However, one spherically symmetric case where further simplification
occurs will be discussed briefly in the next section.
\bigskip

{\bf{5.2\ \ Fluid Sphere of Constant Proper Classical Density}}

Consider a fluid sphere which at the classical level has constant
proper density. Let us suppose that, in addition to the classical
fluid, only massless conformally invariant free fields, such as
the photon and massless neutrino, are present.

The classical interior solution for a fluid sphere of uniform
proper density was found by Schwarzschild in 1916.$^{22}$
It is known that the Weyl tensor of this metric is zero, so that
it is conformally flat.

Because this spacetime is static and has no event horizons, we may suppose
that the the quantum fields are in a well-defined vacuum state.
It has been shown that for conformally
invariant massless free fields in conformally flat spacetimes, the vacuum
stress-energy tensor is determined by the trace anomaly.$^{23,28}$
The vacuum stress-energy tensor of these fields is a linear combination
of $\hbar\ {}^{(1)} H_{ab}$ and $\hbar\ {}^{(3)} H_{ab}$.
We will suppose that there is no additional Casimir energy contribution
in this spacetime.

Therefore, the state-dependent part of the vacuum
expectation value of the quantum stress tensor is zero for massless
conformally invariant free fields propagating on this interior metric.
The only effect of the quantum fields in their vacuum state is to
give rise, through the conformal trace anomaly, to the $\alpha_1$
and $\alpha_3$ state-independent correction terms in the Einstein
equations. Thus, in the interior of the fluid sphere, one has the
{\it complete} equations which must be integrated.

The classical interior Schwarzschild solution has the form
of Eq.\ \SphericalMetric\  with (for $r < R$)
$$A_{cl}(r) = (1 - 2GMr^2/R^3)^{-1},
\eqn\intSchwA$$
and
$$B_{cl}(r) = {1\over 4}
\left[
  3(1 - 2GM/R)^{1/2}-(1 - 2GMr^2/R^3)^{1/2}
\right].
\eqn\intSchwB$$
The classical energy-momentum tensor corresponding to this solution
is that of Eq.\ \FluidStressT , with
constant proper density,
$$\rho = {3M\over 4\pi R^3}.
\eqn\intSchwDensity$$
The pressure $p$ is
$$p(r) = {3M\over 4\pi R^3}
\left[
{
  {(1 - 2GM/R)^{1/2}-(1 - 2GMr^2/R^3)^{1/2}}\over
  {(1 - 2GMr^2/R^3)^{1/2}-3(1 - 2GM/R)^{1/2}}
}
\right].
\eqn\intSchwPressure$$
For $r > R$, this interior metric joins with the classical
Schwarzschild exterior solution of mass $M$, and zero
density and pressure.
It is known that of all stable fluid spheres having a given mass $M$
and radius $R$, the Schwarzschild uniform density sphere has the
smallest central pressure.
For the pressure not to become infinite
somewhere inside the object, it is necessary that $GM < (4/9) R$.
This means that the radius of the static fluid sphere must be larger
than the corresponding Schwarzschild black hole radius.
Quantum corrections may possibly change the relationship between these
two radii for sufficiently small fluid spheres.

 From the previous section, we have the expressions for ${}^{(1)} H_{ab}$
and ${}^{(3)} H_{ab}$ in terms of the classical fluid
$\rho$ and $p$. These are given for the constant density sphere by
Eq.\ \intSchwDensity\  and Eq.\ \intSchwPressure .
The resulting semiclassical Einstein equations are of second order
and are ready for numerical or analytic solution. We will carry this
further in a later paper.
\vfil\eject

\chapternumber=6
\makechapterlabel
\equanumber=0

\centerline{\bigbf{6.  Conclusion}}

We have considered first order semiclassical quantum corrections to  a
variety of classical solutions to the Einstein gravitational field
equations. We have used perturbative constraints to obtain the reduced
semiclassical Einstein equations for Friedman-Robertson-Walker
cosmologies,  for
Friedmann-Lema\^\i tre cosmologies,  for the gravitational fields of
static spherically symmetric fluid bodies, and for the general,
conformally flat metric in terms of its conformal factor.   The reduced
equations we obtained do not contain higher than second derivatives, and
do not exhibit runaway solutions or instabilites of the original fourth
order equations. They have the same physical content as the fourth order
equations, but yield only physically relevant solutions. Analytic and
numerical solutions to these semiclassical equations were found in the
cosmological cases. Although in most cases the semiclassical corrections
play only a small role far from the Planck scale, there are some examples
in which semiclassical quantum corrections cause significant deviation, or
even qualitatively different behavior, from the classical solution.

In the case of spatially flat, radiation-dominated Friedman-Robertson-Walker
solutions, the corrections either strengthen or weaken the singular behavior
at early times, in a regime where the perturbative corrections are valid
(the perturbative validity does break down,
however, before the time of the classical singularity itself can be reached).
The corrections at late times become vanishingly small.
In the Friedmann-Lema\^\i tre ``bounce'' or ``turn-around'' solutions,
quantum corrections can cause classical and semiclassical models which
have the same initial conditions to have significantly different radii at
late times. In the case of the ``maximal hesitation'' Einstein universe,
the semiclassical corrections can cause large deviations and even
qualtitatively different behavior from the corresponding classical
solution.

For fluid spheres, we have given the explicit first order quantum
corrections  as reduced field equations
for the general case, and have discussed the constant
classical density fluid in further detail. In future work, we intend to
study solutions of these equations. For small fluid spheres the
corrections may significantly alter fundamental relations, such as the
classical theorem which requires the radius of a static sphere of fluid to
be larger than the radius of the Schwarzschild black hole having the same
mass.

\bigskip
\centerline{\bigbf{Acknowledgements}}

We thank the National Science Foundation for support of this work under
grant number PHY-9105935. Much of the algebraic and numerical work was done
using {\it Mathematica}$^{25}$ and {\it MathTensor}$^{26}$.

\vfil\eject

\centerline{\bigbf{Appendix A}}

\noindent {\bf{Integration Factor}}

In several instances in this paper, the semiclassical solutions are calculated
iteratively, from an equation of motion of the form
$$\dot a(t)=f_0(a(t))+\hbar f_1(a(t))+O(\hbar^2), \eqno{({\rm A}1)}$$
and an ansatz of
$$a=a_0+\hbar a_1+O(\hbar^2). \eqno{({\rm A}2)}$$
Inserting Eq.\ (A2) into Eq.\ (A1) and expanding in powers of $\hbar$ produces
$$\dot a_0+\hbar\dot a_1=f_0(a_0)+\hbar a_1f'_0(a_0)+\hbar
f_1(a_0)+O(\hbar^2),$$
which leads to the series of equations
$$\eqalign{h^0: &\qquad \dot a_0(t) = f_0(a_0(t))\cr
h^1: &\qquad \dot a_1(t)=a_1(t)f'_0(a_0(t))+f_1(a_0(t)).\cr}$$
The first equation is typically a non-linear equation which might be solved in
a variety of ways.  The second is a first order {\it{linear}} inhomogeneous
equation in $a_1(t)$ (once the classical solution $a_0$ has been determined),
for
which a general solution can be always found in the following form:
$$a_1(t) = {1\over{c\mu (t)}} + {1\over{\mu (t)}} \int^t dt'\;
\mu(t')f_1(a_0(t')) , \eqno{({\rm A}3)}$$
where $c$ is an arbitrary constant of integration and $\mu (t)$ is an
integrating
factor given by
$$\mu (t) = \exp \left( -\int^t dt'\; f'_0(a_0(t'))\right) . \eqno{({\rm
A}4)}$$
Because Eq.\ (A1) has no explicit dependence on $t$, we know that there is a
one
parameter family of solutions, $a_{t_0}(t)=a(t-t_0)$, to Eq.\ (A1),
parametrized by the
initial time $t_0$.  The freedom to choose the constant $c$ in Eq.\ (A3) must
correspond to
the freedom to change this initial time by $t_0\rightarrow\tau =t_0+\hbar t_1$.
By making this shift, and expanding in powers of $\hbar$, we can determine the
integrating factor without the need of integrating Eq.\ (A4) explicitly:
$$a_0(t-t_0)\rightarrow a_0(t-t_0-\hbar t_1)=a_0(t-t_0)-\hbar t_1\dot
a_0(t-t_0)
. \eqno{({\rm A}5)}$$
Comparing this to Eq.\ (A3) reveals that $[c\mu (t)]^{-1}=-t_1\dot a_0(t-t_0)$,
permitting us to rewrite Eq.\ (A3) as
$$a_1(t)=\dot a_0(t)\int^t dt' {f_1(a_0(t'))\over{\dot a_0(t')}} -t_1\dot
a_0(t)+O(\hbar ), \eqno{({\rm A}6)}$$
where $t_1$ is an arbitrary parameter with dimensions of time, and the shift of
initial time in Eq.\ (A5) induces only a higher order $(O(\hbar ))$ change in
the
integrand in  Eq.\ (A6).
\vfil\eject

\centerline{\bigbf{Appendix B}}

\noindent {\bf{Uniqueness of Perturbative Solutions}}

For a general perturbative field equation of the form
$$F_0(q,\dot q,\ddot q)+\epsilon F_1(q,\dot q,\ddot q,{\buildrel ...\over
q},q^{iv}) +\ldots =O(\epsilon^{n+1}) , \eqno{({\rm B}1)}$$
where $\epsilon$ is the formal perturbative expansion parameter ($\epsilon
=\hbar$ for semiclassical gravity) and $q$ represents all the configuration
space
variables, there is some ambiguity in the way a perturbative solution,
$$q=q_0(t)+\epsilon q_1(t)+\ldots +\epsilon^nq_n(t)+O(\epsilon^{n+1}),
\eqno{({\rm B}2)}$$
may be expanded in the same expansion parameter.  For example, defining $q_0$
as
the quantity that satisfies the lowest order field equation,
$$F_0(q_0,\dot q_0,\ddot q_0)=O(\epsilon ) \eqno{({\rm B}3)}$$
does not unambiguously determine $q_0$, because we can shift by any quantity
$O(\epsilon )$, i.e.\ $q_0\rightarrow q_0+\epsilon\delta q$, and the new
quantity
will still satisfy Eq.\ (B3), only  requiring an accompanying shift in the
higher
order terms in the expansion of the solution, i.e.\ $q_1\rightarrow q_1-\delta
q$.  Similar ambiguity exists for the field equations of higher order terms
$$\epsilon\left( {\p F_0(q,\dot q, \ddot q)\over{\p q}} q_1 + {\p F_0(q,\dot q,
\ddot q)\over{\p \dot q}} \ddot q_1 \right) +\epsilon F_1(q,\dot q,\ddot q,
{\buildrel ...\over q}, q^{iv})=O(\epsilon^2)$$
$${\it{etc.}} \eqno{({\rm B}3a)}$$
Despite this ambiguity in breaking up the solution into terms that solve the
field equations order by order, there is no ambiguity in the sum of all such
terms.  This can be seen by positing an additional requirement that the
individual terms of the solution be explicitly independent of the perturbative
expansion parameter,
$${\p q_i\over{\p\epsilon}} = O(\epsilon^{n+1-i}).
\quad\quad\quad i=0,\;\ldots \;,n\eqno{({\rm B}4)}$$
That this requirement can always be met can be easily seen as follows.  Instead
of solving Eq.\ (B3), solve the related equation
$$F_0(q_0,\dot q_0,\ddot q_0) = O(\epsilon^{n+1}). \eqno{({\rm B}5)}$$
Any solution to (B5) is also a solution to (B3), but there is no ambiguity to
$O(\epsilon^n)$.  Similarly solve the analogs of Eq.\ (B3a) to the highest
order
allowed
$$\epsilon\left( {\p F_0(q,\dot q, \ddot q)\over{\p q}} q_1 + {\p F_0(q,\dot q,
\ddot q)\over{\p \dot q}} \dot q_1 + {\p f_0(q,\dot q,\ddot q)\over{\p\ddot q}}
\ddot q_1\right) +\epsilon F_1(q,\dot q,\ddot q, {\buildrel ...\over q},
q^{iv})=O(\epsilon^{n+1})$$
$${\it{etc.}} \eqno{({\rm B}6)}$$
This process uniquely defines each of the terms in Eq.\ (B2), and therefore
also
the sum (to $O(\epsilon^n))$.

Despite the ability to fix the expansion in this manner, it is often to our
advantage to use
the freedom to make order-by-order shifts in the terms $q_i$ of the solution,
as
done in Appendix A.  Solutions with different $\epsilon$-dependencies can be
obtained by adding $\epsilon$-dependent terms
$$\eqalign{q_0 &\rightarrow q_0+\epsilon\delta_1q_0 +
\epsilon^2\delta_2q_0+\ldots +\epsilon^n \delta_nq_0\cr
q_1 &\rightarrow q_1-\delta_1q_0 -\ldots -\epsilon^{n-
1}\delta_nq_0+\epsilon\delta_1q_1 +\ldots +\epsilon^{n-1}\delta_{n-1}q_1\cr
q_2 &\rightarrow q_2-\delta_1q_1 -\ldots -\epsilon^{n-
1}\delta_nq_1+\epsilon\delta_1q_2 +\ldots +\epsilon^{n-1}\delta_{n-2}q_2\cr}$$
$${\rm{etc.}} \eqno{({\rm B}7)}$$
where the $\delta_1q_j$ are arbitrary $O(\epsilon^0)$ functions to be chosen at
one's convenience.
\vfil\eject

\centerline{\bigbf{Appendix C}}

\noindent {\bf{Analytic Semiclassical Solution for Turn-around or Bounce
Universe}}

The semiclassical corrections to the turn-around or bounce universe were
calculated numerically in section 3.3.1, but they can also be calculated
analytically.  The classical turn-around solution is a solution to
Eq.\ \EinsteinttReduced, for positive cosmological constant $(\Lambda >0)$,
spatially closed
slicing $(k=1)$, and sufficiently small radiation density:
$$0< {4\Lambda\kappa\rho a^4_0\over 9} <1. \eqno{({\rm C}1)}$$
The classical solution is
$$a^2_{cl} = {1\over 2} {3\over\Lambda} \left( q\cosh\left[
2{\sqrt{{\Lambda\over
3}}} t\right]+1\right) . \eqno{({\rm C}2)}$$
where
$$q\equiv {\sqrt{1-{4\Lambda\kappa\rho a^4_0\over 9}}} , \eqno{({\rm C}3)}$$
and $0<q<1$.

The ansatz $a=a_{cl}+\hbar a_1+O(\hbar^2)$, when inserted into Eq.\
\EinsteinttReduced
\ and expanded in powers of $\hbar$, gives as the order $\hbar$ equation:
$$\eqalign{O(\hbar ) = 2\dot a_1\dot a_{cl} &- 2 {\Lambda\over 3} a_1a_{cl} +
{2\kappa\rho a^4_0\over 3} a_1a_{cl}^{-3}\cr
&- {8\alpha_1\Lambda\kappa\rho a^4_0\over 3} a_{cl}^{-2}+\alpha_3 \left[
{\Lambda\over 3} a^2_{cl}+ {\kappa\rho a^4_0\over 3} a_{cl}^{-2}\right]^2 a^{-
2}_{cl} .\cr} \eqno{({\rm C}4)}$$
This is a first order, linear, inhomogeneous equation in $a_1(t)$, and it may
be
solved by standard methods shown in Appendix A.  The integrations are tedious,
but easily within the grasp of a good symbolic integration software package.
The
result, up to the initial time ambiguity dealt with in Appendix A, is:
$$\eqalign{a_1 &=\alpha_1 {\sqrt{6\Lambda}} (q-q^{-1})\cosh\left[ 2
{\sqrt{{\Lambda\over 3}}} t\right] \left( q\cosh\left[ 2{\sqrt{{\Lambda\over
3}}}
t\right] +1\right)^{-1/2}\cr
&\qquad +\alpha_3 {{\sqrt{3\Lambda}}\over{24}} \Biggl[ {3\sinh \left[
2{\sqrt{{\Lambda\over 3}}} t\right]\over{{\sqrt{1-q^2}}}} \ln \left( {{\sqrt{1-
q^2}}\sinh\left[ 2{\sqrt{{\Lambda\over 3}}} t\right]-\cosh\left[
2{\sqrt{{\Lambda\over 3}}} t\right] - q\over{q\cosh\left[ 2
{\sqrt{{\Lambda\over
3}}} t\right]+1}}\right)\cr
&\qquad +2 {\sqrt{{\Lambda\over 3}}} t\sinh \left[ 2 {\sqrt{{\Lambda\over 3}}}
t\right] + {1\over q} \left( 5q\cosh\left[ 2 {\sqrt{{\Lambda\over 3}}} t\right]
-
1\right) - {1\over q} {1-q^2\over{q\cosh\left[ 2{\sqrt{{\Lambda\over 3}}}
t\right] +1}}\Biggr]\cr
&\qquad\qquad\qquad\qquad\qquad\qquad\qquad\qquad\qquad \times \left(
q\cosh\left[ 2{\sqrt{{\Lambda\over 3}}} t\right] +1\right)^{-1/2}\cr}$$
$$\qquad\qquad\qquad\qquad\qquad +O(\hbar ) . \eqno{({\rm C}5)}$$
The final result for $a(t)$ is given by inserting Eqs.\ (C2) and (C5) into
the perturbative ansatz, $a=a_{cl}+\hbar a_1+O(\hbar^2)$.
%\vfil\eject
\bigskip
\centerline{\bf References}
\smallskip
\item{1.}  See e.g.\ N. D. Birrell and P. C. W. Davies,
{\it Quantum Fields in Curved Space} (Cambridge University Press,
Cambridge, 1982);
S.\ A.\ Fulling, {\it Aspects of Quantum Field Theory in Curved Space-Time}
(Cambridge University Press,Cambridge, 1989); and references in both.
\item{2.}  We use the conventions $c=1$, $\eta_{ab}={\hbox{diag}}(-1,1,1,1)$,
$R^l_{man}=\p_a\Gamma^l_{mn}+$..., and $R^l_{mln}=R_{mn}R$.
\item{3.} M.\ V.\ Fischetti, J.\ B.\ Hartle, and B.\ L.\ Hu, Phys.\ Rev.\ D
{\bf 20}, 1757
(1979); A.\ A.\ Starobinsky, Phys.\ Lett.\ {\bf 91B}, 99 (1980);
 P.\ Anderson, Phys.\ Rev.\ D {\bf 28}, 271 (1983);
 P.\ Anderson, Phys.\ Rev.\ D {\bf 29}, 615 (1984).
\item{4.} G.\ T.\ Horowitz and R.\ M.\ Wald, Phys.\  Rev.\ D {\bf{17}} (1978)
414; G.\ T.\
Horowitz, {\it{ibid}}.\ {\bf{21}} 1445 (1980).
\item{5.} J.\ B.\ Hartle and G.\ T.\ Horowitz, Phys.\ Rev.\ D {\bf{24}} 257
(1981).
\item{6.}  J.\ Z.\ Simon, Phys.\ Rev. D {\bf{43}}, 3308 (1991).
\item{7.} L.\ Bel and H.\ Sirousse-Zia, Phys.\ Rev.\ D {\bf{32}}, 3128 (1985).
\item{8.} R.\ M.\ Wald, Commun.\ Math.\ Phys. {\bf{54}}, 1 (1977); R.\ M.\
Wald, Ann.\
Phys.\ {\bf{110}}, 472 (1978); R.\ M.\ Wald, Phys.\ Rev.\ D {\bf{17}}, 1477
(1978).
\item{9.} P.\ C.\ W.\ Davies, S.\ A.\ Fulling, S.\ M.\ Christensen and T.\ S.\
Bunch, Ann.\
Phys.\ {\bf{109}}, 108 (1977).
\item{10.}  V.\ L.\ Ginzburg, D.\ A.\ Kirzhnits and A.\ A.\ Lyubushin, Zh.\
Eksp.\ Teor.\
Fiz.\ {\bf{60}}, 451 (1971) [Sov.\ Phys.\ JETP {\bf{33}}, 242 (1971)].
\item{11.} P.\ A.\ M.\ Dirac, Proc.\ Roy.\ Soc.\ {\bf{A167}} 148 (1938).  The
non-
relativistic limit (Abraham-Lorentz model) is described in J.\ D.\ Jackson,
{\it
Classical Electrodynamics} (Wiley, New York, 1975).
\item{12.} T.\ L.\ Curtright, G.\ I.\ Ghandour, and C.\ K.\ Zachos, Phys.\
Rev.\ D
{\bf{34}} 3811 (1986); K.\ Maeda and N.\ Turok, Phys.\ Lett.\ B {\bf 202}, 376
(1988);
R.\ Gregory, {\it ibid.}, {\bf 199}, 206 (1988).
\item{13.}  J.\ Z.\ Simon, Phys.\ Rev.\ D {\bf{41}}  3720 (1990).
\item{14.}  J.\ Z.\ Simon, unpublished.
\item{15.} L.\ Parker and D.\ Toms, Phys.\ Rev.\ D {\bf{32}}, 1409 (1985).
\item{16.} J.\ Z.\ Simon, Phys.\ Rev.\ D {\bf{43}}, 3308 (1991); {\bf{45}} 1953
(1992).
\item{17.} H.\ J.\ Bhabha, Phys.\ Rev.\ {\bf{70}}, 759 (1946).
\item{18.} X.\ Jaen, J.\ Llosa, and A.\ Molina, Phys.\ Rev.\ D {\bf{41}} 2302
(1986).
\item{19.} M.\ R.\ Brown and A.\ C.\ Ottewill, Phys.\ Rev.\ D {\bf{31}}, 2514
(1985).
\item{20.} W.-M.\ Suen, Phys.\ Rev.\ Lett.\ {\bf 62}, 2217 (1989);
Phys.\ Rev.\ D {\bf 40}, 315 (1989).
\item{21.} J.\ B.\ Hartle, in {\it Gravitation in Astrophysics (Carg\'ese
1986)},
Proceedings of the Sumer Institute, Carg\'ese, France, 1986, edited by
B.\ Carter and J.\ Hartle, NATO ASI Series B: Physics, Vol.\ 156 (Plenum, New
York, 1987); J.\ P.\ Paz and S.\ Sinha, Phys.\ Rev.\ D {\bf 44}, 1083 (1991).
\item{22.}  K.\ Schwarzschild, Sitzungsperichte Preuss.\ Akad.\ Wiss.\ {\bf
1916}, 424;
S.\ Weinberg, {\it Gravitation and Cosmology} (Wiley, New York, 1972), pp.\
330-335;
C.\ W.\  Misner, K.\ S.\ Thorne, and
J.\ A.\ Wheeler, {\it Gravitation} (Freeman, San Francisco, 1973), pp.\
609-612.
\item{23.} L.\ S.\  Brown and J.\ P.\ Cassidy,
Phys.\ Rev.\ D {\bf16}, 1712 (1977).
\item{24.} V.\ Sahni, H.\ Feldman, and A.\ Stebbins, Ap.\ J.\ {\bf 385}, 1
(1992),
and references therein.
\item{25.} Wolfram Research, Inc., {\it Mathematica}, (Wolfram Research, Inc.,
Champaign, IL, 1992).
\item{26.} L.\ Parker and S.\ M.\ Christensen, {\it MathTensor},
(MathSolutions,
Inc., Chapel Hill, NC, 1992).
\item{27.} L.\ H.\ Ford, Phys.\ Rev.\ D {\bf 11}, 3370 (1975);
{\bf 14}, 3304 (1976);
J.\ S.\ Dowker and R.\ Critchley, J.\ Phys.\ A {\bf 9}, 535 (1976).
\item{28.} L.\ Parker, ``Aspects of Quantum Field Theory in
Curved Space-Time: Effective Action and Energy-Momentum Tensor,'' in
{\it Recent Developments in Gravitation, Carg\`ese 1978}, edited by
M.\ L\'evy and S.\ Deser pp.\ 219-273, see in particular, pp.\ 253-258.

\vfil\eject
\centerline{\bigbf{Figures}}
\bigskip

\noindent Figure 1. Plots of the scale factor, ${\tilde a}$, $\tilde A$,
and ${\tilde a}_0$  as functions of ${\tilde t}$, for zero cosmological
constant and zero spatial curvature.
The top two plots are for ${\tilde \alpha _3 = + 0.01}$, the bottom two
are for ${\tilde \alpha _3 = - 0.01}$, and the central plot,
${\tilde a}_0$ , is for
${\tilde \alpha _3 = 0}$ (classical solution).
${\tilde a}$ and $\tilde A$ are equally legitimate solutions to the
semiclassical equations, differing only at order $O(\hbar^2)$.
\bigskip

\noindent Figure 2. Plot of $a$ as a function of $t$, each in units of
$(4\Lambda_r/3)^{-1/2}$. The lower curve is the classical solution
($v=0$) with $u=0.5$ and $\Lambda=\Lambda_r$, $\kappa=\kappa_r$. The upper
curve includes quantum corrections, with $v=0.4$ and other parameters
unchanged.
\bigskip

\noindent Figure 3. Solutions to the classical maximal hesitation equations.
The upper plot is a maximal hesitation solution which spends an infinite time
near the Einstein static universe, but pulls away and ends in an infinite
inflationary epoch.
The constant solution is the Einstein static universe.
The lower plot, also a maximal hesitation solution, begins at a singularity
and asymptotically approaches the Einstein static universe at late times.
\bigskip

\noindent Figure 4. Solutions of quantum corrections to the maximal
hesitation equations. Two classical
solutions and their semiclassical counterparts are shown for
$\alpha_1\hbar L_0^{-2} = -\alpha_3\hbar L_0^{-2} = 0.0001$.
Both the classical and semiclassical initial singularity solutions begin
at small scale factor, but the classical solution asymptotically
approaches the static Einstein solution while the semiclassical solution
diverges exponentially from the classical solution (departing from the
perturbative regime). The classical and semiclassical late time de Sitter
solutions both agree at late times, but at early times the classical
solution asymptotically nears the static Einstein solution, and the
semiclassical solution diverges exponentially from the classical solution
(departing from the perturbative regime). The two semiclassical solutions
may be smoothly matched, as seen in the insert, resulting in a non-maximal
hesitation solution that is always perturbatively valid.

\vfil
\end